\documentclass[journal,draftcls,onecolumn,12pt,twoside]{IEEEtranTCOM}
\pdfoutput=1
\usepackage{balance}
\usepackage{setspace}
\usepackage{url}
\usepackage{dblfloatfix} 
\usepackage{array}
\usepackage{multirow}
\newcolumntype{P}[1]{>{\centering\arraybackslash}p{#1}}
\newcolumntype{M}[1]{>{\centering\arraybackslash}m{#1}}
\usepackage{tabularx}
\usepackage{calc}
\usepackage{xcolor}
\usepackage{url}
\usepackage{textcomp}
\usepackage{tabularx}
\usepackage{cite}

\normalsize

\ifCLASSINFOpdf
   \usepackage[pdftex]{graphicx}
   \DeclareGraphicsExtensions{.pdf,.jpeg,.png}
\else
  \usepackage[dvips]{graphicx}
  \DeclareGraphicsExtensions{.eps}
\fi

\usepackage{amsmath, amsthm, amssymb}

\makeatletter
\let\cite\relax
\DeclareRobustCommand{\cite}{%
  \let\new@cite@pre\@gobble
  \@ifnextchar[\new@cite{\@citex[]}}
\def\new@cite[#1]{\@ifnextchar[{\new@citea{#1}}{\@citex[#1]}}
\def\new@citea#1{\def\new@cite@pre{#1}\@citex}
\def\@cite#1#2{[{\new@cite@pre\space#1\if\relax\detokenize{#2}\relax\else, #2\fi}]}
\makeatother
\begin{document}
\title{Entropy and Energy Detection-based Spectrum Sensing over $\mathcal{F}$~Composite Fading Channels}
\author{
Seong~Ki~Yoo,~\IEEEmembership{Member,~IEEE,}~Simon~L.~Cotton,~\IEEEmembership{Senior Member,~IEEE,} \\ Paschalis C. Sofotasios,~\IEEEmembership{Senior Member,~IEEE,} Sami~Muhaidat, ~\IEEEmembership{Senior~Member,~IEEE,} Osamah~S.~Badarneh, \IEEEmembership{Member,~IEEE,} and~George~K.~Karagiannidis,~\IEEEmembership{Fellow,~IEEE}

\thanks{S. K. Yoo and S. L. Cotton are with Centre for Wireless Innovation, ECIT Institute, Queen's University Belfast, Belfast BT3 9DT, U.K.  (e-mail: $\{ \rm sk.yoo;  simon.cotton; \}@qub.ac.uk$).} 

\thanks{P. C. Sofotasios is with the Department of Electrical and Computer Engineering, Khalifa University of Science and Technology,
Abu Dhabi 127788, UAE, and also with the Department of Electronics and Communications Engineering, Tampere University of Technology, Tampere 33101, Finland (e-mail: $\rm p.sofotasios@\rm ieee.org$).}

\thanks{S. Muhaidat is with the Department of Electrical and Computer Engineering, Khalifa University of Science and Technology,
Abu Dhabi 127788, UAE, and also with the Institute for Communication
Systems, University of Surrey, Guildford GU2 7XH, U.K. (e-mail: $\rm muhaidat@ieee.org$).} 

\thanks{O. S. Badarneh is with Electrical Engineering Department, Faculty of Engineering, University of Tabuk, Tabuk 71491, Saudi Arabia  (e-mail: $\rm obadarneh@ut.edu.sa$).} 

\thanks{G. K. Karagiannidis is with the Department of Electrical and Computer Engineering, Aristotle University of Thessaloniki, Thessaloniki 54124, Greece (e-mail: $\rm geokarag@auth.gr$).}
}

\markboth{Submitted to IEEE Journal}%
{Submitted to IEEE Journal}

\maketitle
\begin{abstract}
In this paper, we investigate the performance of energy detection-based spectrum sensing over $\mathcal{F}$~composite fading channels. To this end, an analytical expression for the average detection probability  is firstly derived. This expression is then extended to account for collaborative spectrum sensing, square-law selection diversity reception and noise power uncertainty. The corresponding receiver operating characteristics (ROC) are analyzed for different conditions of the average signal-to-noise ratio (SNR), noise power uncertainty, time-bandwidth product, multipath fading, shadowing, number of diversity branches and number of collaborating users. It is shown that the energy detection performance is sensitive to the severity of the multipath fading and amount of shadowing, whereby even small variations in either of these physical phenomena can significantly impact the detection probability. As a figure of merit to evaluate the detection performance, the area under the ROC curve (AUC) is derived and evaluated for different multipath fading and shadowing conditions. Closed-form expressions for the Shannon entropy and cross entropy are also formulated and assessed for different average SNR, multipath fading and shadowing conditions. Then the relationship between the Shannon entropy and ROC/AUC is examined where it is found that the average number of bits required for encoding a signal becomes small (i.e.,~low Shannon entropy) when the detection probability is high or when the AUC is large. The difference between composite and traditional small-scale fading is emphasized by comparing the cross entropy for Rayleigh and Nakagami-$m$ fading. A validation of the analytical results is provided through a careful comparison with the results of some simulations. 
\end{abstract}


\section{Introduction}

\IEEEPARstart{T}{he} detection of unknown signals is an important issue in many areas of wireless communications such as carrier-sense multiple access based networks, radio detection and ranging (RADAR) systems and cognitive radio \cite{1}. Also, it is expected to be useful in numerous emerging wireless technologies, such as in vehicle-to-vehicle communications, as well as  in  Internet-of-Things (IoT) based applications, where numerous devices are expected to perform sensing in order to communicate to each-other or with other systems or networks \cite{R1,R2,R3} - and the references therein. As a result, there have been a number of signal detection techniques proposed in the literature. Among the most common are matched filter detection (MFD), cyclostationary feature detection (CFD) and energy detection (ED) \cite{new1, new2, new3, 2} and the references therein. Compared to the MFD and CDF techniques, ED is quite attractive as it does not require a priori knowledge of the primary signal, i.e., it is a non-coherent detection method. Thus, ED simply measures the received signal energy level over an observation interval and compares it with a pre-determined threshold to determine the presence or absence of the primary signal. Due to its ease of implementation, ED has understandably gained much attention and widespread use \cite{2, 3, 4}. In particular for cognitive radio, ED is commonly used as a spectrum sensing mechanism in order for the secondary users (SUs) to determine whether a primary user (PU) is present or absent in a given frequency band.

Since the effectiveness of ED-based spectrum sensing is greatly impacted by the fading conditions experienced within the operating environment, its performance has been investigated for a number of commonly encountered fading channels \cite{5, 6, 7, 8, 9, R4}. For example, the behavior of ED-based spectrum sensing over traditional fading channels, such as Rayleigh \cite{5, 6}, Rician \cite{5, 6} and Nakagami-$m$ \cite{5, 6, 7, 8, 9}, has been studied in terms of the false alarm probability $(P_f)$ and detection probability $(P_d)$ or equivalently missed-detection probability $(P_{m} = 1 - P_{d})$. While all of the aforementioned studies have provided important contributions to the understanding of the performance of ED-based spectrum sensing over fading channels, they are restricted to multipath fading channels only. However, in practice, the wireless signal may not only undergo multipath fading but also simultaneous shadowing. 

To take into account concurrent multipath fading and shadowing, several composite fading models have been proposed for conventional and emerging communications channels. Accordingly, the performance of ED-based spectrum sensing has also been evaluated over these composite fading channels \cite{10, 11, 12, 13, 14, 15}. For example, in \cite{10, 11, 12}, the detection performance was investigated within the context of lognormal-based composite fading channels. However, due to the intractability of the lognormal distribution, the Rayleigh / lognormal \cite{11} and Nakagami-$m$ / lognormal \cite{12} composite fading models were approximated using the semi-analytic mixture gamma (MG) distribution. Moreover, a comprehensive performance analysis of ED-based spectrum sensing over generalized $K$ $\left(K_{G}\right)$ composite fading channels \cite{13} and gamma-shadowed Rician fading channels \cite{14} has been conducted for both single-branch and diversity reception cases. More recently, the performance of ED-based spectrum sensing over $\kappa$-$\mu$, $\eta$-$\mu$ and $\alpha$-$\mu$ fading channels and their respective generalized composite fading channels, namely $\kappa$-$\mu$ / gamma, $\eta$-$\mu$ / gamma and $\alpha$-$\mu$ / gamma, has been studied in \cite{R4} and \cite{15}, respectively. In the latter, again due to the inherent mathematical complexity of the formulations, an MG distribution was employed to approximate semi-analytically these three composite fading models.

More recently, in \cite{16}, the authors have proposed the use of the Fisher-Snedecor $\mathcal{F}$ distribution to model composite fading channels in which the root mean square (rms) power of a Nakagami-$m$ signal is assumed to be subject to variations induced by an inverse Nakagami-$m$ random variable. In \cite{16}, it was demonstrated that the $\mathcal{F}$ composite fading model provides as good, and in most cases better fit to real-world composite fading channels compared to the $K_G$ composite fading model. Most importantly, when comparing the analytical forms of the key statistical metrics and performance measures, the $\mathcal{F}$ composite fading model shows significantly less complexity than the $K_G$ composite fading model. Motivated by these observations, in this paper, we analyze the performance of ED-based spectrum sensing over $\mathcal{F}$ composite fading channels. Based on the fact that the entropy of the received signal depends on whether the primary signal is present or absent \cite{entropy1}, we also evaluate the Shannon entropy and cross entropy over $\mathcal{F}$ composite fading channels, which provides interesting insights. The main contributions of this paper are summarized as follows:

\begin{enumerate}
	\item We derive a computationally tractable analytic expression for the average detection probability ($\bar{P_d}$) for ED-based spectrum sensing over $\mathcal{F}$ composite fading channels. 

	\item We then extend this to the cases of collaborative spectrum sensing and square-law selection (SLS) diversity to improve the detection performance. 

	\item We analyze the performance of ED-based spectrum sensing over $\mathcal{F}$ composite fading channels using the receiver operating characteristic (ROC) curves. Comprehensive numerical results provide useful insights into the performance of ED over $\mathcal{F}$ composite fading channels for different average signal-to-noise ratio (SNR) levels, time-bandwidth product, multipath fading conditions, shadowing conditions and number of diversity branches and collaborating users. Furthermore, we also investigate the effect of noise power uncertainty on the detection performance. All of these results will be useful in the design of energy-efficient cognitive radio systems for emerging wireless applications. 

	\item We derive a closed-form expression for the area under the ROC curve (AUC) and  evaluate this for different multipath fading and shadowing conditions.   
	
  \item We derive closed-form expressions for the Shannon entropy and cross entropy over $\mathcal{F}$ composite fading channels.  To the best of authors' knowledge, none of the expressions presented in the paper have been previously reported in the literature.
			
	\item The behavior of the Shannon entropy and cross entropy is then evaluated for different conditions of the average SNR levels, multipath fading and shadowing conditions. Most importantly, we provide important insights into the relationship between the Shannon entropy and energy detection performance. 
\end{enumerate}

The remainder of the paper is organized as follows. In Section~II, we briefly review the principle of ED and the statistical characteristics of the $\mathcal{F}$ composite fading model. In Section~III, we present analytical expressions for the  $\bar{P_d}$ over $\mathcal{F}$ composite fading channels for the cases of single user spectrum sensing, collaborative spectrum sensing, SLS diversity reception and noise power uncertainty. Subsequently, a closed-form expression for the AUC is presented in Section~IV. In Section~V, we also provide exact closed-form expressions for the Shannon entropy and cross entropy over $\mathcal{F}$ composite fading channels. Section~VI provides some numerical and simulation results while Section~VII presents some concluding remarks.

\section{Energy Detection and the $\mathcal{F}$ Composite Fading Model}
\subsection{Energy Detection}
\begin{figure}[!t]
\centering
\includegraphics[width=4.5in]{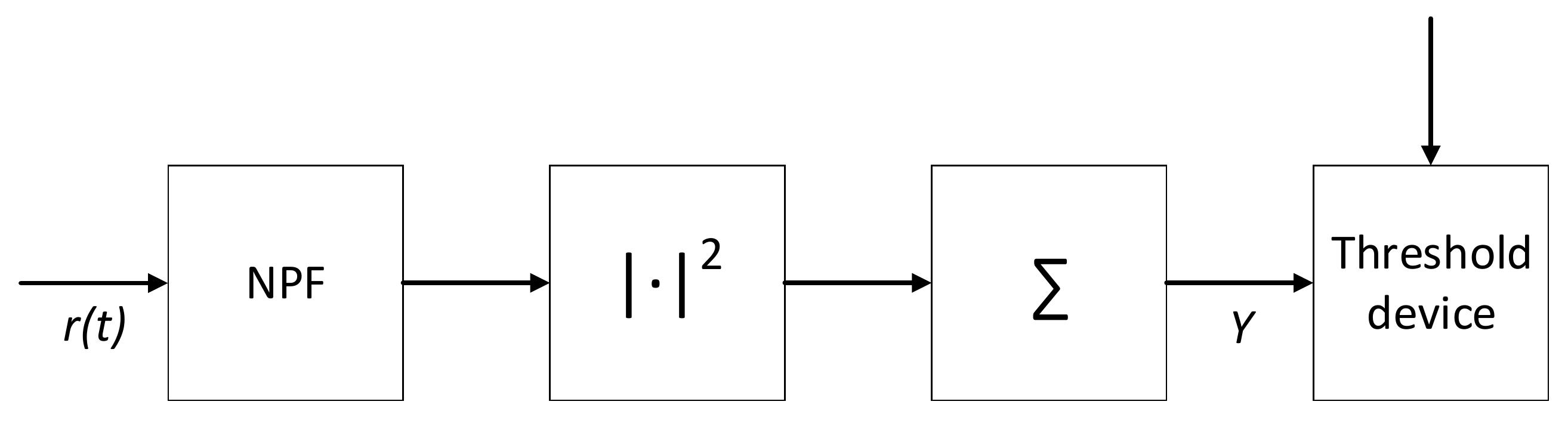}
\caption{System model of energy detection.}
\vspace{-0.4cm}
\label{Fig1}
\end{figure}

The received signal  $r(t)$ at the output of an ED circuit can be described as \cite{5} 

\begin{equation}\label{eqn1}
r(t) = \left\{ {\begin{array}{*{20}{c}}
{\,\,\,\,\,\,\,\,\,\,\,n(t),\,\,\,\,\,\,\,\,\,\,\,\,\,\,\,\,{H_0}}\\ 
{\,\,\,\,\,h(t)\,s(t) + n(t),\,\,\,\,\,\,{H_1}}
\end{array}} \right.
\end{equation}
\noindent where $s(t)$ and $n(t)$ denote the transmitted signal and noise\footnote{For the purposes of modelling, the noise is assumed to be additive white Gaussian noise (AWGN).}  respectively, $h(t)$  represents the complex channel gain and $t$ is the time index. The hypothesis $H_0$ signifies the absence of the signal, conversely the hypothesis $H_1$  represents the presence of the signal. As shown in Fig.~\ref{Fig1}, a typical ED set-up consists of a noise pre filter (NPF), squaring device, integrator and threshold unit. Accordingly, the received signal is first filtered by an ideal bandpass filter within a pre-determined bandwidth ($W$) and then the output of the filter is squared and integrated over an observation interval ($T$) to produce the test statistic ($Y$). The corresponding test statistic is compared with a pre-determined threshold ($\lambda$). 

The test statistic $Y$ can be modeled as a central chi-square random variable where the number of degrees of freedom is equal to twice the time-bandwidth product ($u=TW$), i.e., $2u$ degrees of freedom, under hypothesis $H_0$ \cite{5}. On the other hand, under hypothesis $H_1$, it is modeled as a non-central chi-square random variable with $2u$ degrees of freedom and non-centrality parameter $2\gamma$,  where $\gamma = h^{2} E_{s} / N_0$ is the SNR with $E_s$ and $N_0$ denoting the signal energy and single-sided noise power spectral density respectively. As a result, the corresponding probability density function (PDF) of the test statistic $Y$ can be expressed as follows:
\begin{equation}\label{eqn2}
{f_Y}\left( y \right)\! = \! \left\{ {\begin{array}{*{20}{c}}
{ \displaystyle \frac{{{y^{u - 1}}}}{{{2^u}\Gamma \left( u \right)}}\exp \left( { - \frac{y}{2}} \right),\,\,\,\,\,\,\,\,\,\,\,{H_0}}\\ \\
{\displaystyle \frac{1}{2}{{\left( {\frac{y}{{2\gamma }}} \right)}^{\!\!\!\frac{{u - 1}}{2}}}\!\!\!\!\!\exp \!\left( \!{ - \frac{{2\gamma \! + \!y}}{2}} \!\right)\!{I_{u - 1}}\!\left(\! {\sqrt {2\gamma y} } \right),\,{H_1}}
\end{array}} \right.
\end{equation}
\noindent where $\Gamma[\cdot]$  denotes the gamma function \cite[eq. (8.310.1)]{17} and $I_v(\cdot)$ represents the modified Bessel function of the first kind and order $v$ \cite[eq. (9.6.20)]{18}. Based on the test statistic above, the $P_f$ and $P_d$ of ED over AWGN channels are given as \cite{5} 
\begin{equation}\label{eqn3}
{P_f} = {P_r}\left( {Y > \lambda |{H_0}} \right) = \frac{{\Gamma \left( {u,\,{\lambda  \mathord{\left/
 {\vphantom {\lambda  2}} \right.
 \kern-\nulldelimiterspace} 2}} \right)}}{{\Gamma \left( u \right)}}
\end{equation}
\noindent and
\begin{equation}\label{eqn4}
{P_d} = {P_r}\left( {Y > \lambda |{H_1}} \right) = {Q_u}\left( {\sqrt {2\gamma } ,\,\sqrt \lambda  } \right)
\end{equation}
\noindent where $\Gamma(\cdot,\cdot)$ and $Q_u(\cdot,\cdot)$  represent the upper incomplete gamma function \cite[eq. (8.350.2)]{17} and the generalized Marcum $Q$-function \cite[eq. (1)]{19}, respectively. 

\subsection{The $\mathcal{F}$ Composite Fading Model}
Similar to the physical signal model proposed for the Nakagami-$m$ fading channel, the received signal in an $\mathcal{F}$ composite fading channel is composed of separable clusters of multipath, in which the scattered waves have similar delay times, with the delay spreads of different clusters being relatively large. However, in contrast to the Nakagami-$m$ signal, in an $\mathcal{F}$ composite fading channel, the rms power of the received signal is subject to random variation induced by shadowing. Following this description, the received signal envelope, $R$, can be expressed as 
\setcounter{equation}{4}
\begin{equation}\label{SignalModel}
{R^2} = \sum\limits_{n = 1}^m {{A^2}I_n^2 + {A^2}Q_n^2} 
\end{equation}
\noindent where $m$ represents the number of clusters, $I_n$ and $Q_n$ are independent Gaussian random variables which denote the in-phase and quadrature phase components of the cluster $n$, with $\mathbb{E}[I_n]=\mathbb{E}[Q_n]=0$ and $\mathbb{E}[I^2_n]=\mathbb{E}[Q^2_n]=\sigma^2$, with $\mathbb{E}[\cdot]$  denoting the statistical expectation. In \eqref{SignalModel}, $A$ is a normalized inverse Nakagami-$m$ random variable where $m_s$ is the shape parameter and $\mathbb{E}[A^2]=1$, such that
\begin{equation}\label{InverseNakagami}
{f_A}(\alpha ) = \,\frac{{2{{\left( {{m_s} - 1} \right)}^{{m_s}}}}}{{\Gamma \left( {{m_s}} \right)\,{\alpha ^{2{m_s} + 1}}}}\exp \left( { - \frac{{{m_s} - 1}}{{\,{\alpha ^2}}}} \right).
\end{equation}
\noindent
Using the same approach used in \cite{16}, we can obtain the corresponding PDF of the received signal envelope, $R$, in an $\mathcal{F}$ fading channel as follows
\begin{equation}\label{EnvelopePDF}
{f_R}(r) = \frac{{2\,{m^m}{ \left( {{m_s} - 1} \right)^{{m_s}}\Omega^{{m_s}}}{r^{2m - 1}}}}{{B\left( {m,{m_s}} \right){{\left[ {m{r^2} + \left( {{m_s} - 1} \right)\,\Omega } \right]}^{m + {m_s}}}}}, \quad {m_s} > 1
\end{equation}
\noindent where $B(\cdot,\cdot)$ denotes the beta function \cite[eq. (8.384.1)]{17}. It is worth highlighting that in this paper, we have modified slightly the underlying inverse Nakagami-$m$ PDF from that used in  \cite{16} and subsequently the PDF for the $\mathcal{F}$ composite fading model\footnote{While the PDF given in \cite{16} is completely valid for physical channel characterization, unfortunately we have not been able to determine the parameter range over which the entropy and energy detection performance are computable. On the other hand, the redefined PDF for the $\mathcal{F}$ composite fading model given in \eqref{EnvelopePDF} is well consolidated.}. The form of the PDF in \eqref{EnvelopePDF} is functionally equivalent to the $\mathcal{F}$ distribution.\footnote{Letting $r^2 = x$, $m = d_{1}/2$, $m_s = d_2/2$,  $\Omega = d_2/ (d_2 - 2)$ and performing the required transformation yields the $\mathcal{F}$ distribution, $f_X(x)$, with parameters $d_1$ and $d_2$.} In terms of its physical interpretations, $m$ denotes the fading severity whereas $m_s$ controls the amount of shadowing of the rms signal power. Moreover, $\Omega=\mathbb{E}[r^2]$ represents the mean power. As $m_{s} \rightarrow 0$, the scattered signal component undergoes heavy shadowing conditions. In contrast, as  $m_{s} \rightarrow \infty$,   there exists no shadowing in the channel and therefore it corresponds to a Nakagami-$m$ fading channel. Furthermore, as  $m \rightarrow \infty$ and  $m_{s} \rightarrow \infty$, the $\mathcal{F}$ composite fading model becomes increasingly deterministic, i.e., an AWGN channel.

The corresponding PDF of the instantaneous SNR, $\gamma$, in an $\mathcal{F}$ composite fading channel can be straightforwardly obtained  using the transformation of variable  $\gamma = \bar{\gamma} r^2 / \Omega$, such that
\\
\begin{equation}\label{eqn5}
{f_\gamma }(\gamma ) = \frac{{{m^m}{{(m_s\!-\!1)^{{m_s}}}\,\bar{\gamma}^{{m_s}}}{\gamma ^{m - 1}}}}{{B\left( {m,{m_s}} \right){{\left[ {m\gamma  + (m_s-1)\,\bar{\gamma} } \,\right]}^{m + {m_s}}}}}
\end{equation}
\noindent
where $\bar{\gamma}=\mathbb{E}[\gamma]$ is the average SNR. 

\section{Energy Detection over $\mathcal{F}$ Composite Fading Channels}
\subsection{Single User Spectrum Sensing}

When the signal undergoes fading, the average false alarm probability of ED-based spectrum sensing does not change as $P_f$  is independent of the SNR fading statistics, while the $\bar{P_d}$  of ED can be obtained by averaging over the corresponding SNR fading statistics as follows
\setcounter{equation}{8}
\begin{equation}\label{eqn6}
{\bar P_d} = \int_0^\infty  {{P_d}\,\,{f_\gamma }\left( \gamma  \right)} \,\rm d\gamma.
\end{equation}
\noindent
To this end, the  $\bar{P_d}$ of ED-based spectrum sensing over $\mathcal{F}$ composite fading channels can be obtained by substituting \eqref{eqn4} and \eqref{eqn5} into \eqref{eqn6}, such that 
\begin{equation}\label{eqn7}
{\bar P_d} \!=\! \int_0^\infty \!\!\!\! {{Q_u}\!\left(\! {\sqrt {2\gamma } ,\sqrt \lambda  } \right)\!\frac{{{m^m}{{(m_s\!-\!1)^{{m_s}}}\,\bar{\gamma}^{{m_s}}}{\gamma ^{m - 1}}}}{\!{B\!\left( {m,{m_s}} \right){{\!\left[ {m\gamma \! +\! {(m_s\!-\!1)}\,\bar \gamma } \,\right]}^{m + {m_s}}}}}} \rm d\gamma.
\end{equation}
\noindent
Recognizing that the generalized Marcum $Q$-function in \eqref{eqn7} can be equivalently expressed as \cite[eq. (29)]{20}, namely
\begin{equation}\label{eqn8}
{Q_u}\left( {\sqrt {2\gamma } ,\,\sqrt \lambda  } \right) = \exp \left( { - \gamma } \right)\sum\limits_{n = 0}^\infty  {\frac{{{\gamma ^n}\,\Gamma \left( {n + u,\,{\lambda  \mathord{\left/
 {\vphantom {\lambda  2}} \right.
 \kern-\nulldelimiterspace} 2}} \right)}}{{\Gamma \left( {n + 1} \right)\Gamma \left( {n + u} \right)}}} 
\end{equation}
\noindent
then substituting \eqref{eqn8}  into \eqref{eqn7}, the  $\bar{P_d}$ of ED-based spectrum sensing over $\mathcal{F}$  composite fading channels can be equivalently rewritten as
\begin{equation}\label{eqn9}
{\bar P_d} = \displaystyle \frac{{{m^m}{{(m_s\!-\!1)^{{m_s}}}\,\bar{\gamma}^{{m_s}}}}}{{B\left( {m,\,{m_s}} \right)}}\sum\limits_{n = 0}^\infty  {\frac{{\Gamma \left( {n + u,\,{\lambda  \mathord{\left/
 {\vphantom {\lambda  2}} \right.
 \kern-\nulldelimiterspace} 2}} \right)}}{{\Gamma \left( {n + 1} \right)\Gamma \left( {n + u} \right)}}} \int_0^\infty \!\!\! {\,\frac{{{\gamma ^{n + m - 1}}\exp \left( { - \gamma } \right)}}{{{{\left[ \, {m\,\gamma  + {(m_s\!-\!1)}\,\bar \gamma } \,\right]}^{m + {m_s}}}}}} \,\rm d\gamma.
\end{equation}
\noindent
With the aid of \cite[eq. (3.383.5)]{17} and making use of the generalized Laguerre polynomials \cite[eq. (07.20.03.0018.01)]{21}, the $\bar{P_d}$ can be expressed as 
\\
\begin{equation}\label{eqn10}
\begin{array}{l}
\!{{\bar P}_d} \!=\! \displaystyle  \frac{1}{{B\!\left( {m,{m_s}} \right)}}\! \sum\limits_{n = 0}^\infty \! {\frac{{\Gamma \!\left( {n\! + \!u,\,{\lambda  / 2}} \right)} \Gamma \!\left( {n - {m_s}} \right) }{{\Gamma \! \left( {n \!+ \!1} \right)\Gamma \!\left( {n \!+ \!u} \right)}}} \!\! \left[ \!{{{\left( {\frac{{{(m_s\!-\!1)}\,\bar \gamma }}{m}} \right)}^{\!\!{m_s}}} \!\!\!\! {}_1{F_1}\! \left( \! {m\! +\! {m_s};{m_s}\! - \!n \!+\! 1;\frac{{{(m_s\!-\!1)}\,\bar \gamma }}{m}} \right)} \right.\\ [14pt]
\qquad \qquad \qquad \,\,\, \displaystyle  + \left. {{{\left( {\frac{{{(m_s-1)}\,\bar \gamma }}{m}} \right)}^n} \frac{B\left( {n + m,\,{m_s} - n} \right)}{\Gamma \left( {n - {m_s}} \right)} {}_1{F_1}\left(\! {n \!+ \!m;n \!-\! {m_s} \!+\! 1;\frac{{{(m_s\!-\!1)}\,\bar \gamma }}{m}} \right)} \right]
\end{array}
\end{equation}
\noindent where ${}_1{F_1}\left(\cdot;\cdot;\cdot \right)$ represents the Kummer confluent hypergeometric function \cite[eq. (9.210.1)]{17}. It is worth noting that \cite[eq. (3.383.5)]{17} is only valid when $m+m_{s}$  is not a positive integer number, i.e., $m+m_{s} \neq \mathbb{N}$.   Nonetheless, this potential singularity can be straightforwardly circumvented by introducing an infinitely small perturbation term that can be added to $m+m_{s}$,  if required.  Furthermore, with the aid of \cite[eq. (07.20.16.0006.01)]{21}, \eqref{eqn10} can be rewritten as follows 
\begin{equation}\label{eqn10_v2}
{{\bar P}_d} = \displaystyle \frac{(m_s\!-\!1)^{m_s} {\bar \gamma}^{m_s}}{{B\!\left( {m,{m_s}} \right)} \, m^{m_s}} \sum\limits_{n = 0}^\infty  {\frac{{\Gamma \left( {n + u,\,{\lambda  \mathord{\left/
 {\vphantom {\lambda  2}} \right.
 \kern-\nulldelimiterspace} 2}} \right)}\Gamma \left( n + m \right)}{{\Gamma \left( {n + 1} \right)\Gamma \left( {n + u} \right)}}} \, {U} \! \left( {m\! + \!{m_s};{m_s} \!- \!n \!+ \!1;\frac{{{(m_s\!-\!1)}\,\bar \gamma }}{m}} \right)
\vspace{0.01cm}
\end{equation}
\noindent where $U\!\left(\cdot; \cdot; \cdot \right)$ denotes the confluent hypergeometric function of the second kind.

It is noted that the infinite series representation in \eqref{eqn10_v2} is convergent and only few terms are required in practice of its truncation. However, in the the analysis of digital communications over fading channels, it is essential to determine the exact number of truncation terms to guarantee target performance or  quality of service requirements. Based on this, we derive an upper bound for the truncation error of \eqref{eqn10_v2}, which can be computed straightforwardly because it is expressed in closed-form in terms of known and built-in functions. Based on this, the truncation error, $\mathcal{T}$, for the infinite series in \eqref{eqn10_v2} if it is truncated after  $T_0 - 1$ terms, is given as
\begin{equation}\label{truncation_1}
\mathcal{T} = \displaystyle \!\sum\limits_{n = T_0}^\infty  {\! \frac{{\Gamma \left( {n + u,\,{\lambda  / 2}} \right)} \, \Gamma \left( n + m \right)}{{\Gamma  \left( {n + 1} \right) \Gamma  \left( {n + u} \right)}}} {U} \! \left( {m + {m_s};\,\,{m_s} - n + 1\,;\,\,\frac{{{(m_s-1)}\,\bar \gamma }}{m}} \right).
\end{equation}
\noindent Since the confluent hypergeometric function of the second kind is monotonically decreasing with respect to $n$, $\mathcal{T}$ can be bounded as
\begin{equation}\label{truncation_2}
\mathcal{T} \leq \displaystyle   {U} \! \left( {m + {m_s};\,\,{m_s} - T_0 + 1\,;\,\,\frac{{{(m_s-1)}\,\bar \gamma }}{m}} \right)  \sum\limits_{n = T_0}^\infty  {\frac{{\Gamma \left( {n + u,\,{\lambda  / 2}} \right)}\Gamma \left( n + m \right)}{{\Gamma \left( {n + 1} \right)\Gamma \left( {n + u} \right)}}}. 
\end{equation}
\noindent With the aid of the monotonicity properties of the upper incomplete gamma function, $\Gamma(a, x) < \Gamma(a,0) = \Gamma(a)$, the above expression can be upper bounded as follows
\begin{equation}\label{truncation_3}
 \mathcal{T} < \displaystyle   {U} \! \left(\! {m \!+\! {m_s};{m_s} \!-\! T_0 \!+\! 1;\frac{{{(m_s\!-\!1)}\,\bar \gamma }}{m}} \right) \! \sum\limits_{n = T_0}^\infty \! {\frac{\Gamma \!\left( n \!+\! m \right)}{{\Gamma \!\left( {n \!+\! 1} \right)}}}. 
\end{equation}
\noindent Since we add up strictly positive terms, the summation above can be rewritten as
\begin{equation}\label{truncation_4}
\sum\limits_{n = T_0}^\infty  {\frac{\Gamma \left( n + m \right)}{{\Gamma \left( {n + 1} \right)}}} \leq \sum\limits_{n = 0}^\infty  {\frac{\Gamma \left( n + m \right)}{{\Gamma \left( {n + 1} \right)}}}. 
\end{equation}
\noindent To this effect and by also recalling the Pochhammer symbol identities, it follows that
\begin{equation}\label{truncation_5}
\mathcal{T}  <  \displaystyle   {U} \! \left(\! {m \!+\! {m_s};{m_s} \!-\! T_0 \!+\! 1;\frac{{{(m_s\!-\!1)}\,\bar \gamma }}{m}} \right)\!\! \sum\limits_{n = 0}^\infty  \!{\frac{(m)_n \, \Gamma(m)}{{n!}}}. 
\end{equation}
\noindent It is evident that the above infinite series representations can be expressed in closed-form as follows
\begin{equation}\label{truncation_6}
\mathcal{T} < \displaystyle   {U} \! \left(\! {m \!+\! {m_s};{m_s} \!-\! T_0 \!+\! 1;\frac{{{(m_s\!-\!1)}\,\bar \gamma }}{m}} \right)\! \Gamma(m) {}_1{F_0}(m;\, ;\, 1)
\end{equation}
\noindent where ${}_1{F_0}(\cdot;\, \cdot;\, \cdot)$ denotes the generalized hypergeometric function.

\subsection{Collaborative Spectrum Sensing}
The detection performance of ED-based spectrum sensing can be significantly improved using collaborative spectrum sensing \cite{23, 24} which exploits the spatial diversity among SUs (i.e., sharing their sensing information). For simplicity, we assume that all $N$ collaborative SUs experience independent and identically distributed (i.i.d.) fading and employ the same decision rule (i.e., the same threshold). For the OR-rule or equivalently 1-out-of-$n$ rule, the final decision is made when at least one SU shares a local decision. In this case, the collaborative detection probability $( {P_d^{\,\rm OR}})$ and false alarm probability $( {P_f^{\,\rm OR}})$ under AWGN can be written as follows
\begin{equation}\label{eqn11}
P_d^{\,\rm OR} = 1 - {\left( {1 - {P_d}} \right)^{N}}
\end{equation}
\noindent and
\begin{equation}\label{eqn12}
P_f^{\,\rm OR} = 1 - {\left( {1 - {P_f}} \right)^{N}}.
\end{equation}
\noindent
On the other hand, for the AND-rule, the final decision is made when all SUs share their local decision. In this case, the ${P_d^{\,\rm AND}}$ and $ {P_f^{\,\rm AND}}$ under AWGN can be written as follows
\begin{equation}\label{eqn13}
P_d^{\,\rm AND} = { {P_d}^{N}}
\end{equation}
\noindent and 
\begin{equation}\label{eqn14}
P_f^{\,\rm AND} = {{P_f}^{N}}.
\end{equation}
\noindent
Based on this, the average detection probability of ED system over $\mathcal{F}$ composite fading channels with $N$ collaborative SUs can be obtained by substituting \eqref{eqn10_v2} into \eqref{eqn11} and \eqref{eqn13} for the OR- and AND-rule respectively, yielding the following analytical representations
\begin{equation}\label{eqn15}
 \bar P_d^{\,\rm OR} \!=\! \displaystyle 1 - \left[ 1\! - \! \frac{(m_s\!-\!1)^{m_s} {\bar \gamma}^{m_s}}{{B\!\left( {m,{m_s}} \right)} \, m^{m_s}} \!\sum\limits_{n = 0}^\infty \! {\frac{{\Gamma \!\left( {n \!+ \!u,\,{\lambda / 2}} \right)}\Gamma \!\left( n \!+\! m \right)}{{\Gamma \!\left( {n \!+\! 1} \right)\Gamma \!\left( {n \!+\! u} \right)}}}  {U} \! \left( \! {m \!+\! {m_s};{m_s} \!- \!n \!+ \!1;\frac{{{(m_s\!-\!1)}\,\bar \gamma }}{m}} \right) \! \right]^{N}
\end{equation}
\noindent and 
\begin{equation}\label{eqn16}
\! \bar P_d^{\,\rm AND} \!=\! \displaystyle \left[ \frac{(m_s\!-\!1)^{m_s} {\bar \gamma}^{m_s}}{{B\!\left( {m,{m_s}} \right)}  m^{m_s}} \!\! \sum\limits_{n = 0}^\infty  \! {\frac{{\Gamma \!\left( {n \!+ \!u,\,{\lambda / 2}} \right)}\Gamma \!\left( n \!+\! m \right)}{{\Gamma \!\left( {n \!+ \!1} \right)\Gamma \!\left( {n \!+ \!u} \right)}}}  {U} \! \left( \! {m \!+\! {m_s};{m_s} \!-\! n \!+\! 1 ;\frac{{{(m_s\!-\!1)}\,\bar \gamma }}{m}} \right) \! \right]^{N}\!\!\!.
\end{equation}

\begin{figure}[!t]
\centering
\includegraphics[width=5.5in]{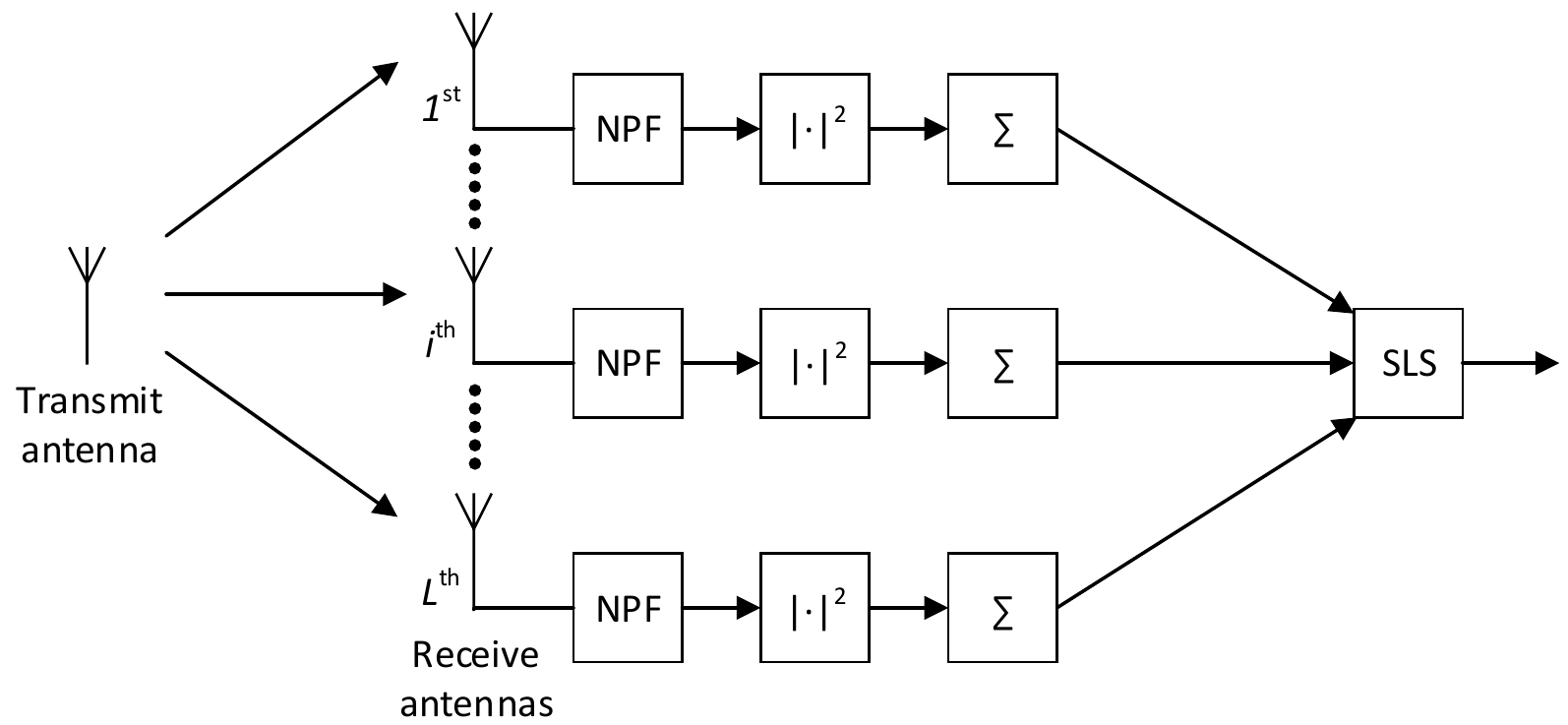}
\caption{System model of energy detection for an $L$-branch SLS diversity scheme.}
\label{Fig2}
\end{figure}

\subsection{Square-Law Selection Diversity Reception}
Using diversity reception techniques is one of the most well-known methods which can be used to mitigate the deleterious effects of fading in wireless communication systems. Among other competing schemes,  SLS diversity reception is efficient and highly regarded due to its simplicity. As shown in Fig. 2, in an SLS scheme, the energy detection process is performed before combining. Consequently, an SLS scheme selects the branch with the highest resultant test statistic, i.e., ${Y^{\rm SLS}} = \max \left\{ {{Y_1},{Y_1}, \ldots {Y_L}} \right\}$  \cite{9}. Under hypotheses $H_0$, the false alarm probability for an SLS scheme $(P_f^{\rm SLS})$ over AWGN channels can be determined as follows
\begin{equation}\label{eqn17}
P_f^{\rm  SLS} = 1 - {\left[ {1 - \frac{{\Gamma \left( {u,\,{\lambda  \mathord{\left/
 {\vphantom {\lambda  2}} \right.
 \kern-\nulldelimiterspace} 2}} \right)}}{{\Gamma \left( u \right)}}} \right]^L}
\end{equation}
\noindent
where $L$ represents the number of diversity branches. On the contrary, under hypothesis $H_1$, the detection probability for an $L$-branch SLS scheme $(P_d^{\rm SLS})$ over AWGN channels can be expressed as 
\begin{equation}\label{eqn18}
P_d^{\rm SLS} = 1 - \prod\limits_{i = 1}^L {\left[ {1 - {Q_u}\left( {\sqrt {2{\gamma _i}} ,\,\sqrt \lambda  } \right)} \right]}.
\end{equation}
\noindent
Consequently, for an $L$-branch SLS system operating over i.i.d. $\mathcal{F}$ composite fading channels, the average detection probability, $P_d^{\rm SLS}$, can be obtained as
\begin{equation}\label{eqn19}
\begin{split}
\bar P_d^{\rm SLS} &=   1 - \prod\limits_{i = 1}^L \!{\int_0^\infty \!\! {\left[ {1 - {Q_u}\!\left( \!{\sqrt {2{\gamma _i}} ,\sqrt \lambda  } \right)} \right]\!{f_{{\gamma _i}}}\left( {{\gamma _i}} \right)} } \, \rm {d}{\gamma _i} \\  
&  = 1 \!-\! \prod\limits_{i = 1}^L {\!\left[ {\int_0^\infty  \!\!\!\!\!{{f_{{\gamma _i}}}\!\left( {{\gamma _i}} \right)} {\rm d}{\gamma _i}\! -\!\! \int_0^\infty \!\!\!\!\! {{Q_u}\!\!\left( \!\!{\sqrt {2{\gamma _i}},\! \sqrt \lambda  } \right)\!{f_{{\gamma _i}}}\!\!\left( {{\gamma _i}} \right)} \, \rm {d}{\gamma _i}} \right]}  \\
& = 1 - \prod\limits_{i = 1}^L {\left[ {1 - {{\bar P}_d}\left( {{\gamma _i}} \right)} \right]} .
\end{split}
\end{equation}
\noindent
By substituting \eqref{eqn10_v2} into \eqref{eqn19}, an analytical expression for $P_d^{\rm SLS}$ is obtained, such that
\begin{equation}\label{eqn20}
\begin{split}
\!\bar P_d^{\rm SLS}  = & 1 - \prod\limits_{i = 1}^L \! \left[ {1 - \frac{(m_{{s_i}}\!\!-\!1)^{m_{s_i}} \, {{\bar \gamma }_i}^{m_{s_i}}}{{B\!\left( {{m_i},{m_{{s_i}}}} \right)}{m_i}^{\! m_{s_i}}}}  \sum\limits_{n = 0}^\infty  \frac{{\Gamma \!\left( {n \!+ \!u,{\lambda / 2}} \right)} \Gamma \!\left( n \!+ \!m_i \right)}{\Gamma \! \left( {n \!+ \!1} \right)\Gamma \!\left( {n \!+ \!u} \right)} \right. \\
& \qquad \qquad \qquad \qquad \qquad \qquad \left.  \times \, U\!\left(\! {m_i}\! + \!{m_{{s_i}}};{m_{s_i}} \!\!-\! n \!+ \!1;\frac{{{(m_{{s_i}}\!\!-\!1)}{{\bar \gamma }_i}}}{{{m_i}}}\,\right) \right].
\end{split}
\end{equation}

\subsection{Noise Power Uncertainty}
In all of the previous cases considered, the detection probability has been grounded on the assumption that the noise power is accurately known. However, in practice, noise power varies with time and location, an effect which is often referred to as the noise power uncertainty \cite{25}. Clearly, any change in the noise power will affect the detection performance, with the main sources of this uncertainty including the non-linearity and the thermal noise of the components in the receiver and environmental noise caused by the transmissions of other wireless users \cite{26, R5, R6}. Therefore, in practice, it is very difficult to obtain a precise knowledge of the noise power.

Assuming that the uncertainty in the noise power estimation can be characterized by the term $\beta$ (which is expressed in decibels), the noise power uncertainty in energy detection can be appropriately modeled as existing in the range $\left[ {{\sigma _W}/\alpha ,\alpha {\sigma _W}} \right]$ where $\sigma _W$  denotes the nominal noise power and $\alpha = 10^{\beta/10} > 1$ quantifies the size of the uncertainty \cite{25}. Therefore, when the noise power is overestimated as ${\bar \sigma _W} = \alpha {\sigma _W}$  (i.e., for the worst case scenario), the detection probability can be obtained as \cite{27}
\begin{equation}\label{eqn21}
P_d^{\rm NU} = {Q_u}\left( {\sqrt {2\gamma } ,\,\sqrt {{\alpha ^2}\lambda } } \right).
\end{equation}
\noindent 
Hence, to evaluate the performance of \eqref{eqn21} over $\mathcal{F}$ composite fading channels, the average detection probability, $\bar P_d^{\rm NU}$, can be directly obtained by scaling the detection threshold with the noise uncertainty, i.e., $\lambda$  is replaced by  $\alpha^2 \lambda$ in \eqref{eqn10_v2}.

\section{Average Area Under the ROC Curve for $\mathcal{F}$ Composite Fading Channels}

The ROC curve is usually employed to evaluate the detection performance. However, for multiple energy detectors, it is difficult to visually compare their performance based on their ROC curves. Following the Area Theorem presented in \cite{28}, the AUC can be used as an alternative measure of the detection capability, where the AUC is simply defined as the area covered by the ROC curve. This metric represents the probability that choosing the correct decision at the detector is more likely than choosing the incorrect decision \cite{29, R7}. As the threshold used in the energy detection varies from $\infty$ to 0 the AUC varies from 0.5 (poor performance) to 1 (good performance).

\vspace{-0.4cm}
\subsection{AUC for the Instantaneous SNR}
Let $A\left(\gamma\right)$ denote the AUC which is a function of instantaneous SNR value $\gamma$. For the ROC curve of 𝑃$P_d$𝑃𝑑 versus $P_f$,  $A\left(\gamma\right)$ can be evaluated as \cite{30}
\begin{equation}\label{eqn22}
A\left( \gamma  \right) = \int_0^1 {{P_d}\left( {\gamma ,\lambda } \right)\,{\rm d}{P_f}\left( \lambda  \right)}.
\end{equation}
\noindent
As both $P_d\left( \gamma, \lambda \right)$ and $P_f\left( \lambda \right)$ are functions of the threshold $\lambda$, we can use the threshold averaging method \cite{31} to calculate the AUC. When the value of $P_f\left( \lambda \right)$ varies from 0 to 1 ($0 \rightarrow 1$), it is equivalent to $\lambda$𝜆 ranging from $\infty$ to 0 ($\infty \rightarrow 0$). Consequently, \eqref{eqn22} can be rewritten as
\begin{equation}\label{eqn23}
A\left( \gamma  \right) =  - \int_0^\infty  {{P_d}\left( {\gamma ,\lambda } \right)\,\frac{{\partial {P_f}\left( \lambda  \right)}}{{\partial \lambda }}} \,{\rm d}\lambda
\end{equation}
\noindent
where $ \frac{{\partial {P_f}\left( \lambda  \right)}}{{\partial \lambda }}$ denotes the partial derivative of $P_f$𝑃𝑓 with respect to $\lambda$, which is obtained from \eqref{eqn3}
\begin{equation}\label{eqn24}
\frac{{\partial {P_f}\left( \lambda  \right)}}{{\partial \lambda }} =  - \frac{{{\lambda ^{u - 1}}}}{{{2^u}\Gamma \left( u \right)}}\exp \left( { - \frac{\lambda }{2}} \right).
\end{equation}
\noindent
By substituting \eqref{eqn4} and \eqref{eqn24} into \eqref{eqn23},   $A\left(\gamma\right)$ can be expressed as
\begin{equation}\label{eqn25}
A\left( \gamma  \right) = \displaystyle \sum\limits_{n = 0}^\infty  {\frac{{{\gamma ^n}\,\exp \left( { - \gamma } \right)}}{{{2^u}\Gamma \left( u \right)\Gamma \left( {n + 1} \right)\Gamma \left( {n + u} \right)}}} \int_0^\infty  {{\lambda ^{u - 1}}\,\exp \left( { - \frac{\lambda }{2}} \right)\Gamma \left( {n + u,\,{\lambda / 2}} \right)} \,{\rm d}\lambda
\end{equation}
\noindent
which with the aid of \cite[eq. (12)]{new_AUC},  it can now be obtained as
\begin{equation}\label{eqn26}
A\left( \gamma  \right)  = 1 - \sum\limits_{l = 0}^{u-1} \sum\limits_{i = 0}^{l} \binom{l + u - 1}{l-i} \frac{\gamma^i }{i! \, 2^{\,l+u+i}} \exp \left( -\frac{\gamma}{2} \,\right)
\end{equation}
\noindent where $\binom{a}{b}$ represents the binomial coefficient. 

\subsection{Average AUC for $\mathcal{F}$ Composite Fading Channels}
The corresponding average AUC $(\bar{A})$ for $\mathcal{F}$ composite fading channels can be evaluated through averaging \eqref{eqn26} by the corresponding SNR fading statistics, such that \cite{30}
\begin{equation}\label{eqn27}
\bar A = \int_0^\infty  {A\left( \gamma  \right)\,{f_\gamma }\left( \gamma  \right)} \,{\rm d}\gamma.
\end{equation}
\noindent
Substituting \eqref{eqn5} and \eqref{eqn26} into \eqref{eqn27},  the average AUC can be expressed as
\begin{equation}\label{eqn28}
\bar A \!= \displaystyle 1 - \sum\limits_{l = 0}^{u-1} \sum\limits_{i = 0}^{l} \binom{l + u - 1}{l-i} \frac{ m^m (m_s-1)^{m_s} \bar{\gamma}^{m_s} }{i! \, 2^{l+u+i} B(m, m_s)} \int_0^\infty  {\frac{{{\gamma ^{m \!+\! i \!-\! 1}}\exp \left(-\frac{\gamma}{2} \right)}}{{{{\left[ {m\,\gamma  + {(m_s\!-\!1)}\bar \gamma } \right]}^{m + {m_s}}}}}} \,{\rm d}\gamma .
\end{equation}
\noindent
Since the integral in \eqref{eqn28} is the same form as that given in \eqref{eqn9}, $\bar A$  can be similarly obtained with the aid of \cite[eq. (3.383.5)]{17} and \cite[eq. (07.20.16.0006.01)]{21}, such that
\begin{equation}\label{eqn29}
\bar A \!= \displaystyle 1 \!- \sum\limits_{l = 0}^{u-1} \sum\limits_{i = 0}^{l} \binom{l\! +\! u \!- \!1}{l\!-\!i} \frac{  (m_s\!-\!1)^{m_s} \bar{\gamma}^{m_s} \Gamma(m+i) }{i! \, 2^{l+u+m_s} m^{m_s} B(m, m_s)} U\!\left(\! {m \!+\! {m_s};{m_s} \!- \!i \!+ \!1;\frac{{{(m_s\!-\!1)}\,\bar \gamma }}{2m}} \right)
\end{equation}
\noindent which is expressed by an exact closed-form expression that involves known functions that are built-in in popular software packages such as Maple, Matlab and Mathematica. 

\section{Entropy for $\mathcal{F}$ Composite Fading Channels}
\subsection{Shannon Entropy}
It is recalled that the Shannon entropy denotes the amount of information contained in a signal and indicates the average number of bits required for encoding this information. For continuous random variables with PDF $p_X(x)$, it is given by $H(p) = -\int_0^\infty  {p_X(x) \log_2\left(p_X(x)\right)} \, {\rm d}x $ \cite{entropy2}. Thus, for the case of $\mathcal{F}$ composite fading channels, it can be expressed by substituting \eqref{eqn5} into $p_X(x)$, such that
\begin{equation}\label{eqn30}
H(p) = \displaystyle - \! \int_0^\infty \!\!\! \frac{{{m^m}{{(m_s\!-\!1)^{{m_s}}}\,\bar{\gamma}^{{m_s}}}{\gamma ^{m - 1}}}}{{B\left( {m,{m_s}} \right){{\left[ {m\gamma  + (m_s\!-\!1)\,\bar{\gamma} } \,\right]}^{m + {m_s}}}}} \log_2 \!\!\left(\!\frac{{{m^m}{{(m_s\!-\!1)^{{m_s}}}\,\bar{\gamma}^{{m_s}}}{\gamma ^{m - 1}}}}{{B\left( {m,{m_s}} \right){{\left[ {m\gamma  + (m_s\!-\!1)\,\bar{\gamma} } \,\right]}^{m + {m_s}}}}} \right)\,{\rm d}\gamma.
\end{equation}
\noindent
Using the logarithmic identities and after some algebraic manipulations, \eqref{eqn30} can be re-written as
\begin{equation}\label{eqn31}
\begin{split}
\!\!\! H(p) &=  - {\log _2}\!\! \left[ {\frac{{{m^m}{{(m_s\!-\!1)^{{m_s}}}\,\bar{\gamma}^{{m_s}}}}}{{B\left( {m,{m_s}} \right)}}} \right] \!-\! \frac{{m^m}{{(m_s\!-\!1)^{{m_s}}}\,\bar{\gamma}^{{m_s}}}}{{B\left( {m,{m_s}} \right)\ln (2)}} \\ 
& \,\, \times \left[ \int_0^\infty  \frac{{{\gamma ^{m - 1}}\ln \left( {{\gamma ^{m - 1}}} \right)}}{{{{\left[ {m\gamma  + \left( {{m_s}\! - 1} \right)\,\bar \gamma } \, \right]}^{m + {m_s}}}}}{\rm d}\gamma  - \int_0^\infty  {\frac{{{\gamma ^{m - 1}}\ln \left( {{{\left[ {m\gamma  + \left( {{m_s}\! - \!1} \right)\,\bar \gamma } \right]}^{m + {m_s}}}} \right)}}{{{{\left[ {m\gamma  + \left( {{m_s}\! -\! 1} \right)\,\bar \gamma } \, \right]}^{m + {m_s}}}}}{\rm d}\gamma  }  \, \right].
\end{split}
\end{equation}
\noindent Performing a simple transformation of variables and applying \cite[eq. (4.293.14)]{17} along with some algebraic manipulation, \eqref{eqn31} can be expressed in closed-form, such that
\begin{equation}\label{eqn32}
\!H(p) \!=\!  \displaystyle   \frac{{(m \!+ \!{m_s})\psi(m\!+\!m_s) \!-\! (m\!-\!1)\psi(m) \!- \!(m_s \!+ \!1) \psi(m_s) }}{{\ln (2)}} \! + {\log _2}\!\left[ {\frac{{B\left( {m,{m_s}} \right)}{{{\left( {{m_s}\! - \!1} \right)}}\,{{\bar \gamma }}}}{m}} \right] 
\end{equation}
\noindent where $\psi \left(\cdot\right)$ represents the psi (polygamma) function \cite[eq. (8.360)]{17}.

\subsection{Cross Entropy}
The cross entropy measures the average number of bits required to encode a message when a distribution $p_X(x)$ is replaced by a distribution $q_X(x)$. The cross entropy between two continuous random variables with PDFs $p_X(x)$ and $q_X(x)$ is given by $H(p,q)=-\int_0^\infty  {p_X(x) \log_2\left(q_X(x)\right)} \, {\rm d}x $ \cite{entropy2}. In the present analysis, $p_X(x)$ represents the $\mathcal{F}$ distribution while the Rayleigh and Nakagami-$m$ distributions are considered for $q_X(x)$. To understand what happens when composite fading is not taken into account, the corresponding cross entropy with respect to the Rayleigh and Nakagami-$m$ distributions are respectively given by
\begin{equation}\label{eqn33}
 H_{\rm Ray}(p,q) \!=\! \displaystyle -\!\! \int_0^\infty \!\!\!\!\! \frac{{{m^m}{{(m_s\!-\!1)^{{m_s}}}\,\bar{\gamma}^{{m_s}} }{\gamma ^{m - 1}}}}{{B\left( {m,{m_s}} \right)\! {{\left[ {m\gamma  + (m_s\!-\!1)\,\bar{\gamma} } \,\right]}^{m + {m_s}}}}} \log_2\left( \frac{1}{\bar \gamma_{R}}\exp\left(- \frac{\gamma}{\bar \gamma_{R}}\right) \right)\,{\rm d}\gamma 
\end{equation}
and
\begin{equation}\label{eqn34}
 H_{\rm Nak}(p,q) \!=\! \displaystyle -\!\! \int_0^\infty \!\!\!\!\! \frac{{{m^m}{{(m_s\!-\!1)^{{m_s}}}\,\bar{\gamma}^{{m_s}} }{\gamma ^{m - 1}}}}{{B\left( {m,{m_s}} \right)\! {{\left[ {m\gamma  + (m_s\!-\!1)\,\bar{\gamma} } \,\right]}^{m + {m_s}}}}}  \log_2\left( \frac{\hat{m}^{\hat{m}} {\gamma}^{\hat{m}-1} }{\Gamma(\hat{m}) \, {\bar \gamma_{N}}^{\,\hat{m}}}\exp\left(- \frac{\hat{m}\gamma}{\bar{\gamma}_{N}}\right) \right)\,{\rm d}\gamma 
\end{equation}
\noindent
where ${\bar{\gamma}_{R}}$ is the average SNR of the Rayleigh distribution, $\hat{m}$ and ${\bar{\gamma}_{N}}$ denote the fading severity parameter and average SNR of the Nakagami-$m$ distribution, respectively. In a similar manner to Section V.A, by performing the necessary transformation of variables and applying \cite[eq. (3.194.3)]{17} and \cite[eq. (4.293.14)]{17} along with some algebraic manipulation, \eqref{eqn33} and \eqref{eqn34} can be expressed in closed-form as follows
\begin{equation}\label{eqn35}
H_{\rm Ray}(p,q) \!=\! \displaystyle \log_2\left( {\bar{\gamma_{R}}} \right) + \frac{\bar \gamma}{\ln\left(2\right)\bar \gamma_{R}}
\end{equation}
and
\begin{equation}\label{eqn36}
{H_{\rm Nak}(p,q)} = \frac{\hat{m} \bar{\gamma}}{\ln(2) \bar{\gamma}_{N}} - \log_{2} \left( \frac{\hat{m}^{\hat{m}}}{\Gamma(\hat{m}) \bar{\gamma}^{\hat{m}}_{N} } \right) + \frac{\hat{m}\!-\!1}{\ln(2)} \left[ \ln \left( \frac{m}{(m_s \! - \! 1) \bar{\gamma}} \right) \!-\! \psi \left( m \right) \!+\! \psi \left( {{m_s}} \right) \right]\!.
\end{equation}
\noindent It is noted that (42) and (46)  can be computed straightforwardly since $\psi(\cdot)$  is included as a built-in function in most popular scientific software packages. 

\section{Numerical and Simulation Results}
Capitalizing on the derived analytic results, we next quantify the effects of $\mathcal{F}$ composite fading conditions for different communication scenarios and fading severity conditions.  

\begin{figure}[!t]
\centering
\includegraphics[width=4.5in]{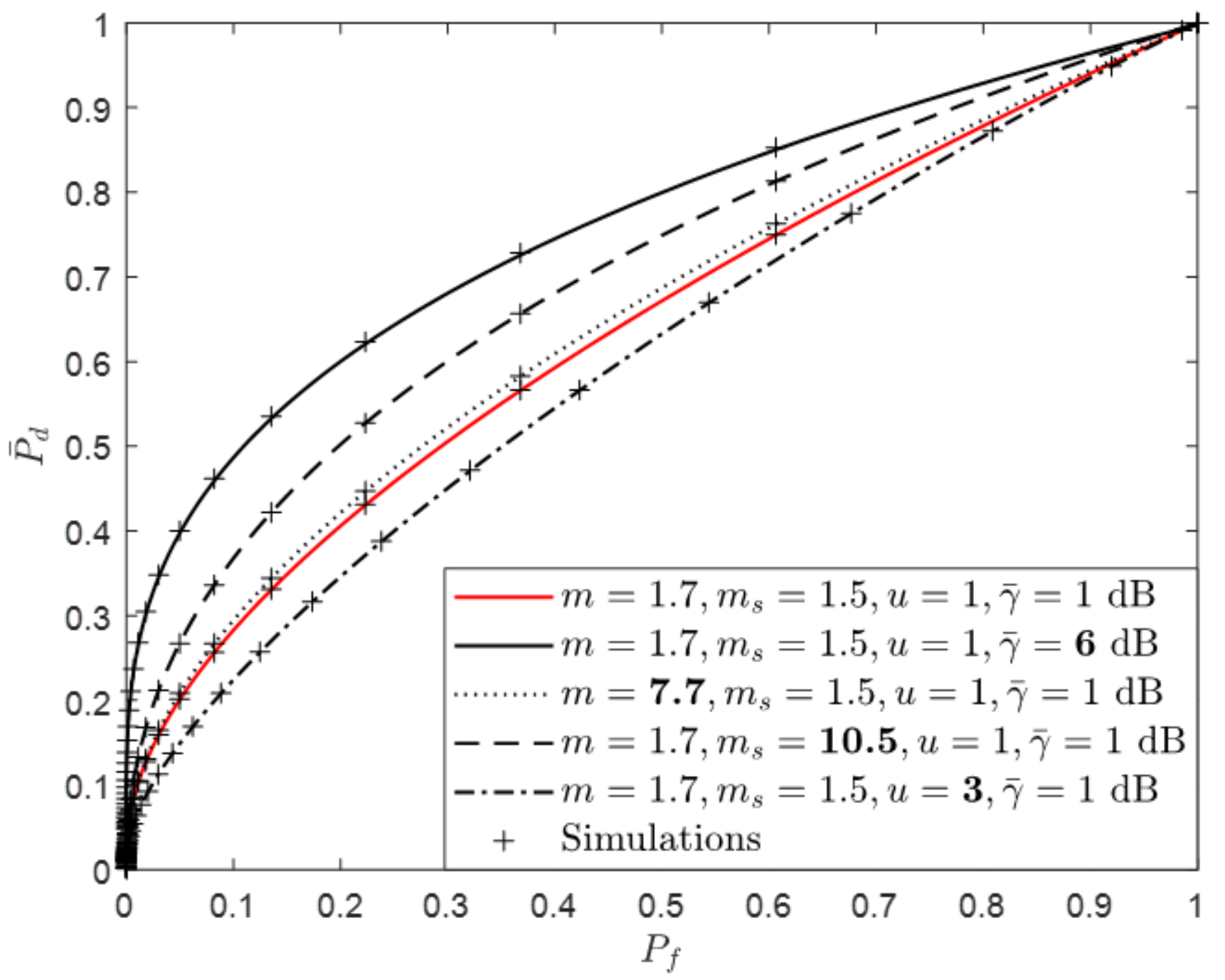}
\vspace{-0.5cm}
\caption{ROC curves for $\mathcal{F}$ composite fading channels considering different $\bar{\gamma}$, $m$, $m_s$ and $u$ values.}
\vspace{-0.7cm}
\label{Fig3}
\end{figure}

\subsection{Energy Detection}

\begin{figure}[!t]
\centering
\includegraphics[width=4.1in]{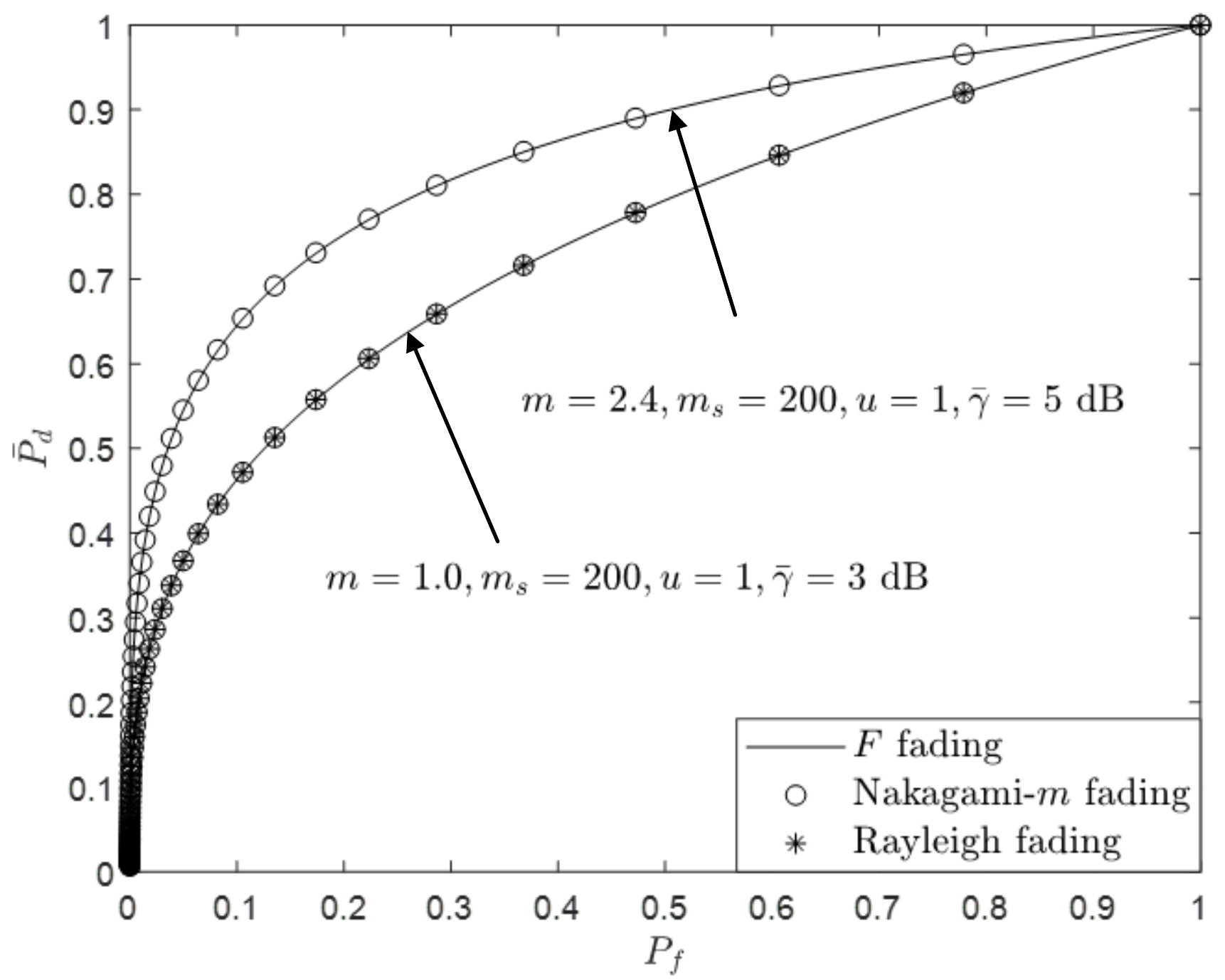}
\vspace{-0.5cm}
\caption{ROC curves for some special cases of the $\mathcal{F}$ composite fading channel: Rayleigh (asterisks) and Nakagami-$m$ (circles).}
\label{Fig4}
\end{figure}

We firstly analyze the performance of ED-based spectrum sensing over $\mathcal{F}$ composite fading channels in terms of the corresponding ROC curves. As an example, Fig. 3 shows the ROC curves for different values of the average SNR ($\bar{\gamma}$), time-bandwidth product ($u$), multipath fading ($m$) and shadowing ($m_s$) parameters. It can be seen that the performance of ED-based spectrum sensing improves when the average SNR increases (higher values of  $\bar{\gamma}$), or when the time-bandwidth product decreases (lower values of $u$), or when the severity of multipath fading and shadowing decreases (higher values of $m$ and $m_s$). It is worth remarking that we have also included the results of some simulations (shown as symbols) in Fig.~\ref{Fig3}, which were performed to validate the derived analytic expressions. Owing to the simplicity of the $\mathcal{F}$ composite fading model, these simulated sequences, each consisting of 100,000 realizations, were straightforwardly generated in MATLAB through the calculation of the ratio of two gamma random variables. In Fig.~\ref{Fig4}, some of the special cases of the ROC curves which coincide with those for the Rayleigh ($m=1$ and $m_s \rightarrow \infty$) and Nakagami-$m$ ($m=m$ and $m_s \rightarrow \infty$) fading channels \cite{5} are illustrated as a further validations and insights.

\begin{figure}[!t]
\centering
\includegraphics[width=4.1in]{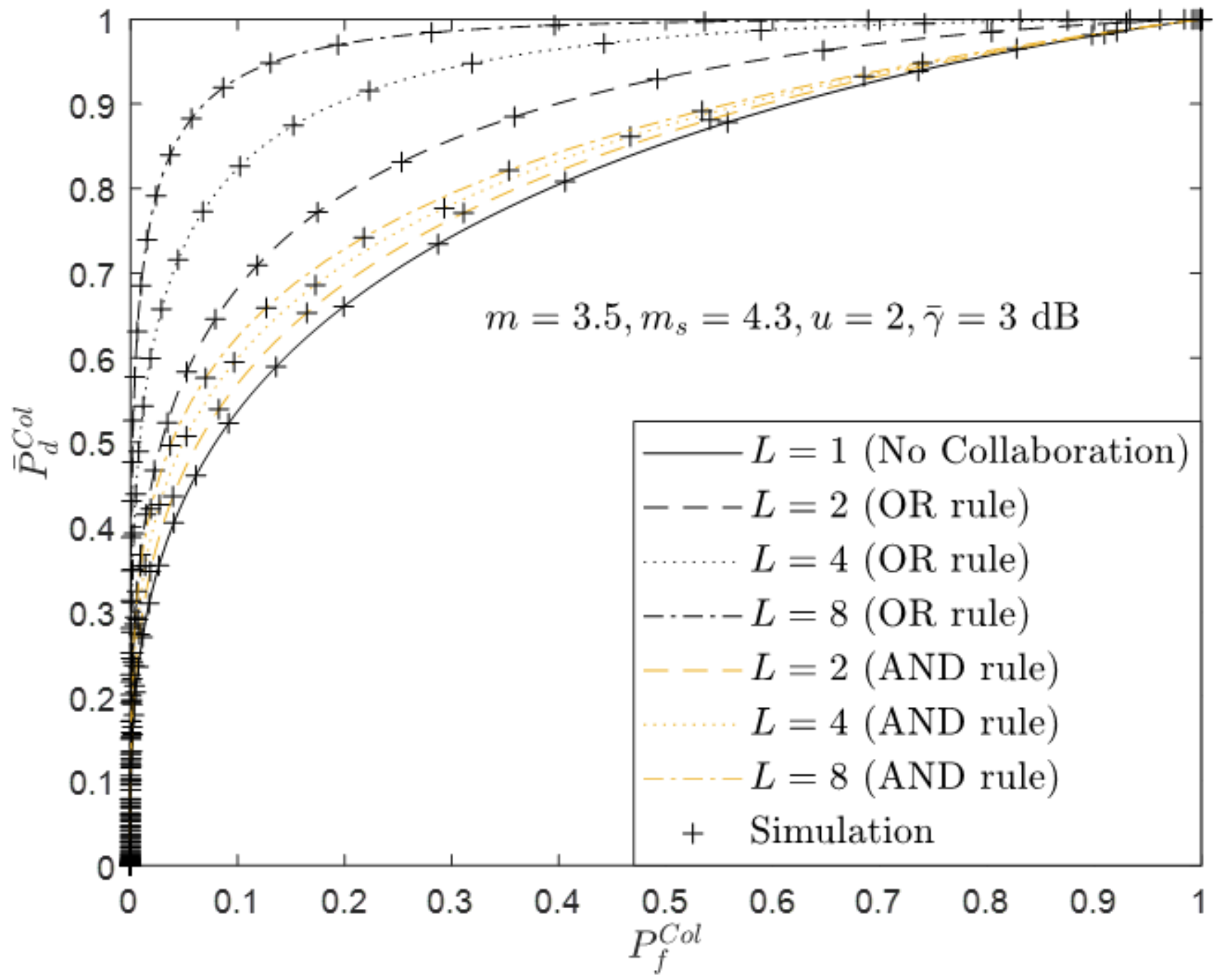}
\vspace{-0.5cm}
\caption{ROC curves for collaborative ED-based spectrum sensing with OR and AND rules over $\mathcal{F}$ composite fading channels, with $m = 3.5$, $m_s = 4.3$, $\bar{\gamma} = 3$ dB, $u = 2$ and $N$ collaborating users with $N $ = 2, 4 and 8. }
\label{Fig5}
\end{figure}

\begin{figure}[!t]
\centering
\includegraphics[width=4.1in]{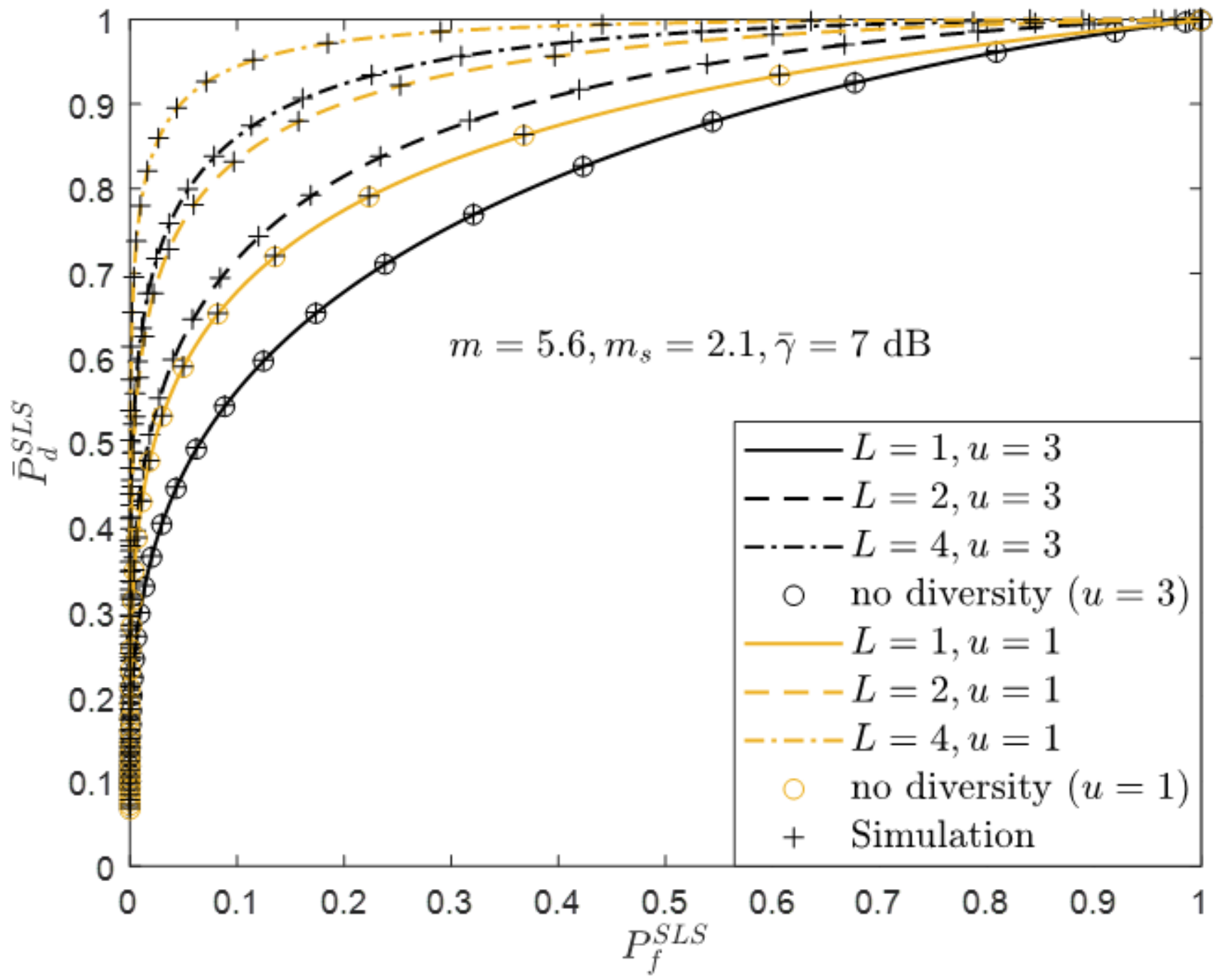}
\vspace{-0.5cm}
\caption{ROC curves for an $L$-branch SLS system with $L$ = 1, 2 and 4 over i.i.d. $\mathcal{F}$ composite fading channels, with $m = 5.6$, $m_s = 1.1$, $\bar{\gamma} = 7$ dB and $u=$\{1, 3\}.}
\vspace{-0.9cm}
\label{Fig6}
\end{figure}

Fig.~\ref{Fig5} illustrates  the ROC curves for collaborative ED-based spectrum sensing with $N = 2, 4, 8$ using the OR and AND rules with $m = 3.5$, $m_s = 4.3$,  $\bar{\gamma} = 3$~dB and $u = 2$. For comparison, the ROC curve for non-collaborative ED-based spectrum sensing (i.e., single user spectrum sensing) is also shown in Fig.~\ref{Fig5}. As expected, the energy detection performance improves as the number of collaborative SUs increases. It is also observed that the OR rule provides a better performance compared to the AND rule. Fig.~\ref{Fig6} shows the detection performance variation with increasing number of diversity branches ($L$) and time-bandwidth product ($u$) for an $L$-branch SLS scheme. It is clear that lower $u$ and higher $L$ provides a better performance. Furthermore, when $L = 1$ the ROC curves for an $L$-branch SLS scheme become equivalent to those for $\mathcal{F}$ composite fading channel. Again, the simulation results provide a perfect match to the analytical results presented in Fig.~\ref{Fig5} and Fig.~\ref{Fig6}.

\begin{figure}[!t]
\centering
\includegraphics[width=4.1in]{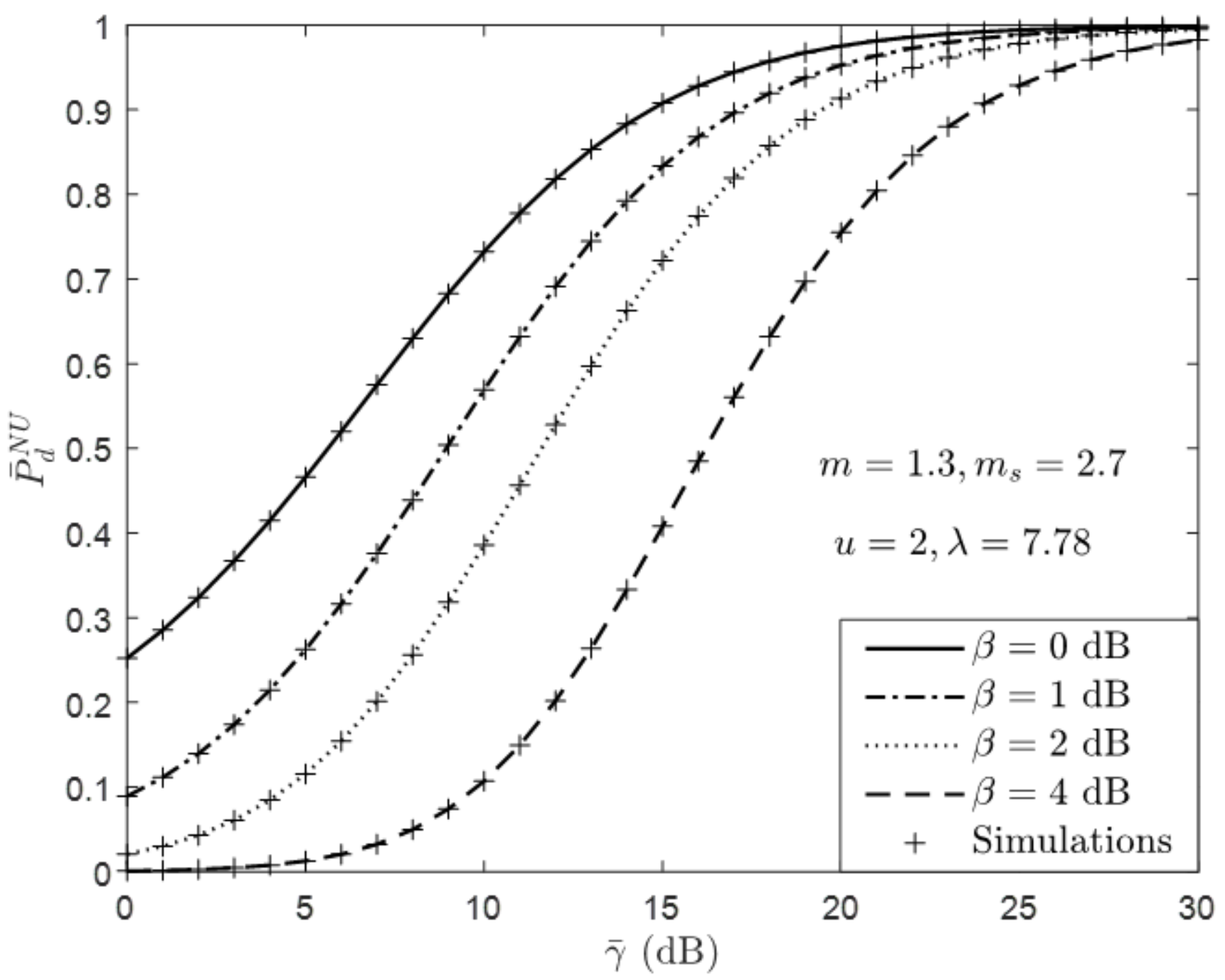}
\vspace{-0.5cm}
\caption{Average detection probability ($\bar P_d^{\rm NU}$) versus average SNR ($\bar{\gamma}$) over $\mathcal{F}$ composite fading channels, with $m = 1.3$, $m_s = 2.7$, $u = 2$ and $\lambda = 7.78$  under different conditions of noise power uncertainty.}
\label{Fig7}
\end{figure}
\begin{figure}[!t]
\centering
\includegraphics[width=4.5in]{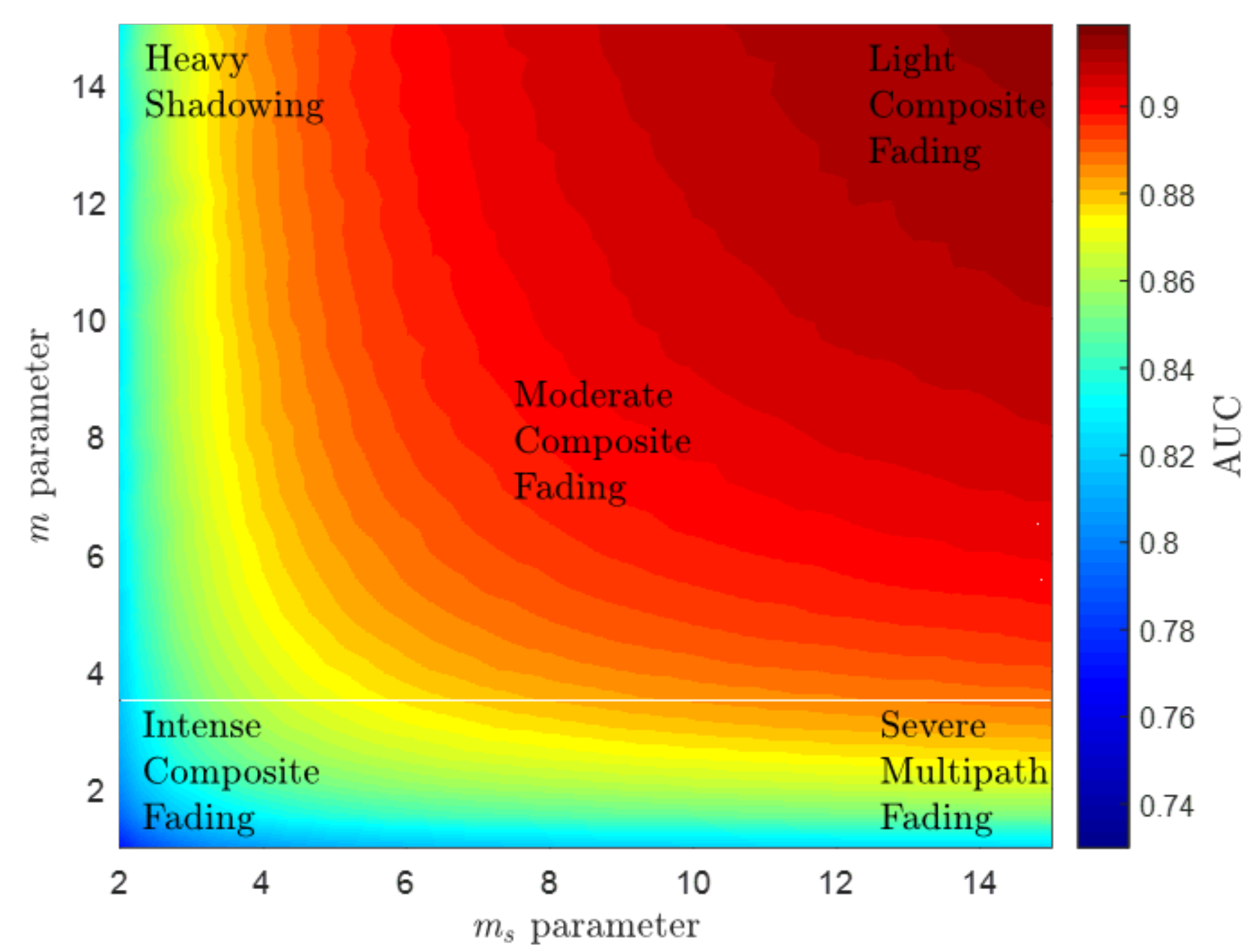}
\vspace{-0.5cm}
\caption{Average AUC in an $\mathcal{F}$ composite fading channel as a function of the multipath fading ($m$) and shadowing ($m_s$) parameters, with $u = 2$ and  $\bar{\gamma} = 2$~dB.}
\label{Fig8}
\end{figure}

Fig.~\ref{Fig7} demonstrates  how the detection performance varies with $\bar{\gamma}$  over $\mathcal{F}$ composite fading channels with $m = 1.3$, $m_s = 2.7$, $u = 2$ and   $\lambda = 7.78$ under a number of different conditions of noise power uncertainty. It is apparent that the detection performance decreases as noise uncertainty increases and the effects of noise uncertainty are non negligible. For example, the value of $\bar P_d^{\rm NU}$ for $\bar{\gamma} = 6$ dB and $\beta = 0$ dB (i.e., perfect noise power estimation) was approximately 0.52 while the value of $\bar P_d^{\rm NU}$ for $\beta = 2$ dB was found to be approximately 0.15. Furthermore, to achieve  $\bar P_d^{\rm NU}= 0.9$, the ED-based spectrum sensing with $\beta = 2$ dB requires an additional 5~dB  compared to when $\beta = 0$ dB. To illustrate both the isolated and combined effects of multipath and shadowing on the AUC for $\mathcal{F}$ composite fading channels, Fig.~\ref{Fig8} shows the estimated AUC values for different multipath fading ($1.0\leq m \leq15$) and shadowing ($2.0\leq m_{s} \leq15$) conditions, with $u = 2$ and  $\bar{\gamma} = 2$~dB. It is clear that smaller values of the AUC (close to 0.5) occurred when the channel was subject to simultaneous heavy shadowing ($m_{s} \rightarrow 2$) and severe multipath fading ($m \rightarrow 1$), i.e., intense composite fading, whereas the higher AUC values (close to 1) appeared when both the multipath and shadowing parameters became large ($m, m_{s} \rightarrow 15$), i.e., light composite fading.

\begin{figure}[!t]
\centering
\includegraphics[width=4.1in]{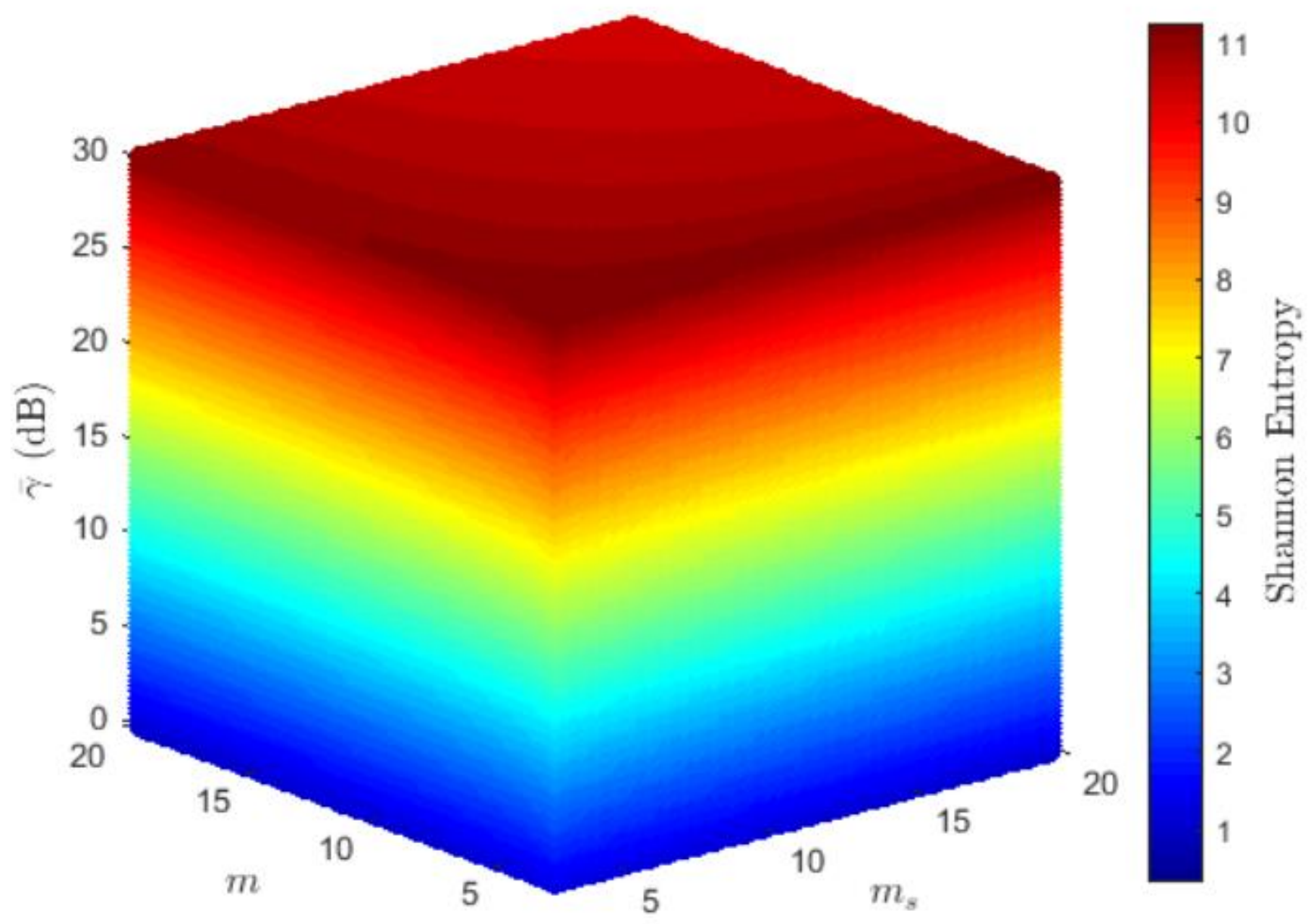}
\vspace{-0.5cm}
\caption{Shannon entropy in an $\mathcal{F}$ composite fading channel as a function of its key parameters: multipath fading ($m$), shadowing ($m_s$) and average SNR ($\bar{\gamma}$).}
\vspace{-0.8cm}
\label{Fig9}
\end{figure}

\subsection{Entropy}
Fig.~\ref{Fig9} shows the estimated Shannon entropy for different multipath fading and shadowing intensities of $\mathcal{F}$ composite fading channels, i.e., $3 \le m \le 10$, $3 \le  m_s \le 10$ and 0 dB $\le \bar{\gamma} \le$ 20~dB. It is obvious that higher values of the Shannon entropy appear at higher $\bar{\gamma}$, lower $m$ and lower~$m_s$. This may indicate that more bits are required to encode the corresponding message when the channel is subject to higher average SNR, severer multipath fading and heavier shadowing. As already shown in Fig.~\ref{Fig3} and Fig.~\ref{Fig8}, the ROC curves and AUC are also highly dependent upon multipath fading and shadowing conditions experienced in $\mathcal{F}$~composite fading channels. Motivated by this, we compare  the behavior of the Shannon entropy and ROC curves.

\begin{figure}[!t]
\centering
\includegraphics[width=3.55in]{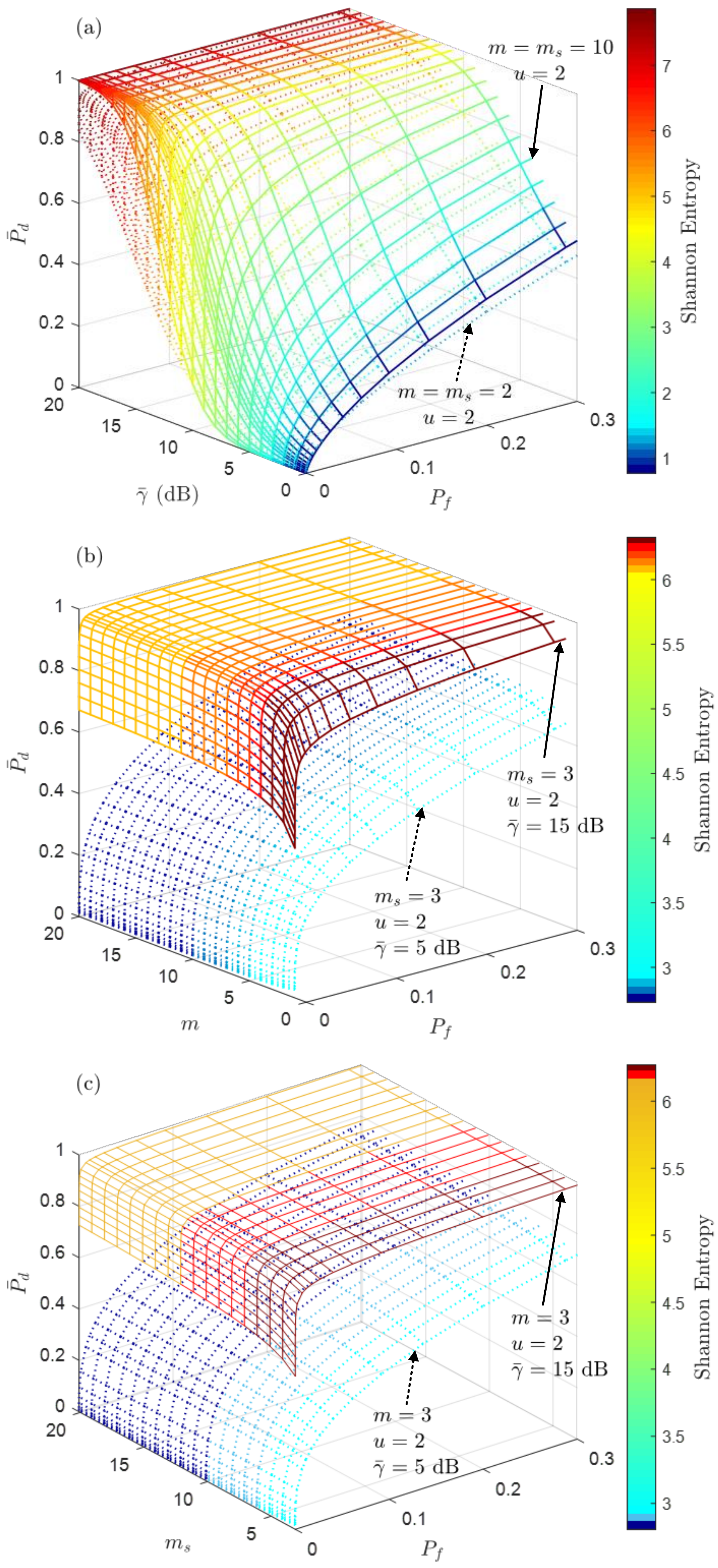}
\vspace{-0.3cm}
\caption{Behavior of the Shannon entropy and ROC curves as a function of (a) average SNR ($\bar{\gamma}$), (b) multipath fading ($m$) and (c) shadowing ($m_s$) parameters, respectively.}
\label{Fig10}
\end{figure}

\begin{figure}[!t]
\centering
\includegraphics[width=4.1in]{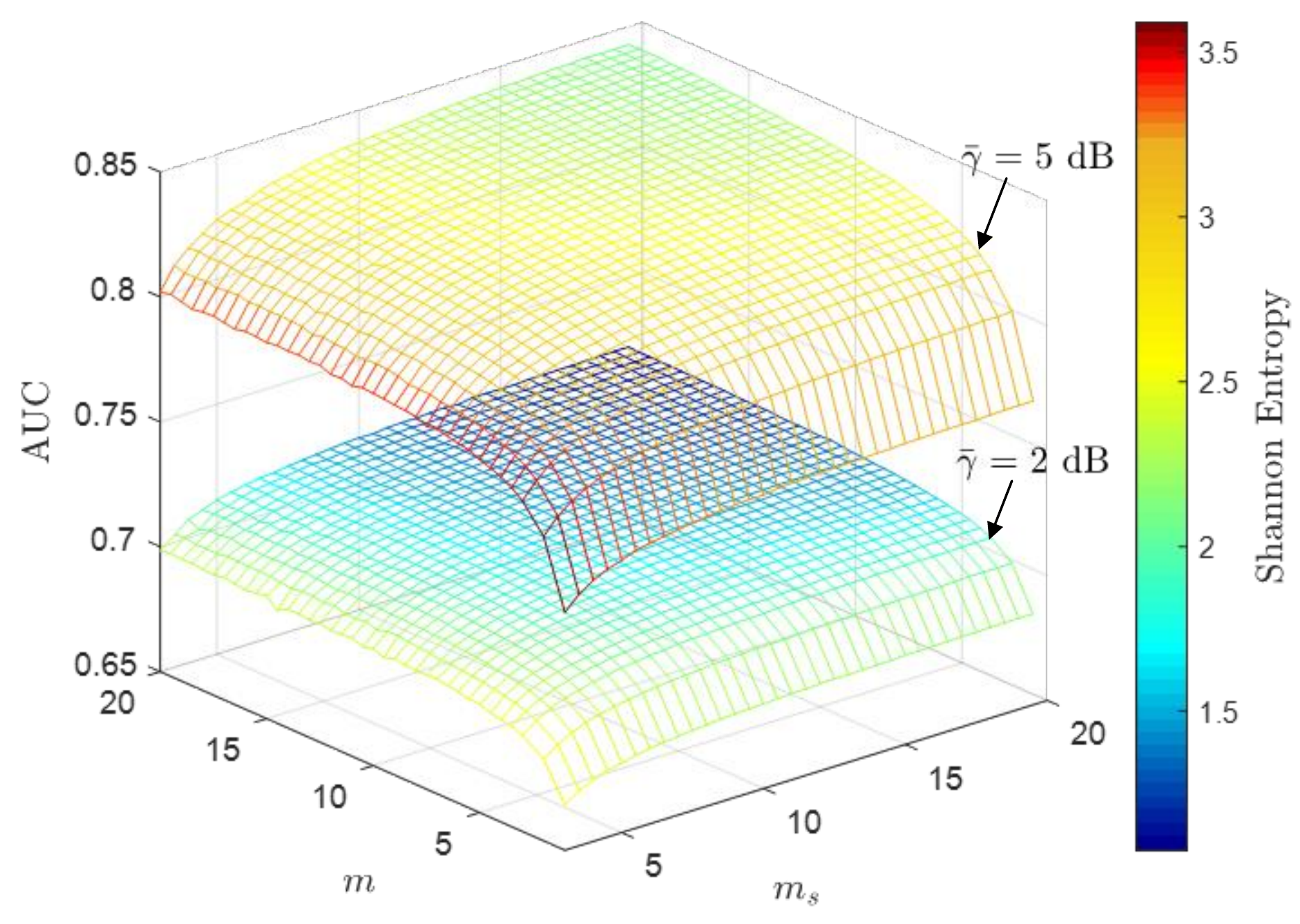}
\vspace{-0.6cm}
\caption{Behavior of the Shannon entropy and AUC as a function of multipath fading ($m$) and shadowing ($m_s$) parameters at $\bar{\gamma}=\{2,5\}$ dB.}
\vspace{-0.8cm}
\label{Fig11}
\end{figure}

Fig.~\ref{Fig10} shows the estimated Shannon entropy and ROC curves as a function of (a) average SNR ($\bar{\gamma}$) with fixed fading parameters $m=m_s=\{2,10\}$ and $u=2$; (b) multipath fading ($m$) with $\bar{\gamma}=\{5,15\}$ dB, $m_s=3$ and $u=2$; (c) shadowing ($m_s$) with $\bar{\gamma}=\{5,15\}$ dB, $m=3$ and $u=2$. It is worth noting that realistic values of $P_f$ in the range 0 to 0.3 (i.e., low false alarm probability) were mainly considered in Fig.~\ref{Fig10}. Similarly, Fig.~\ref{Fig11} compares the behavior of the Shannon entropy and AUC for different values of the multipath fading ($m$) and shadowing ($m_s$) parameters at $\bar{\gamma}=2$ and $5$~dB. It can be easily seen that the value of the Shannon entropy increases when the average SNR increases (higher ~$\bar{\gamma}$), or when the severity of multipath fading increases (lower~$m$), or when the shadowing conditions become heavier (lower~$m_s$). On the other hand, the values of the average detection probability and AUC increase when the average SNR increase (higher~$\bar{\gamma}$), or when the severity of multipath fading decreases (higher~$m$), or when the shadowing conditions become lighter (higher~$m_s$). Consequently, it can be inferred that higher detection capability requires less number of bits for encoding the signal at the same value of $\bar{\gamma}$.

Table~I depicts the estimated Shannon entropy and cross entropy for different values of the fading parameters at  $\bar{\gamma}=5$ and $15$~dB. To obtain the Nakagami-$m$ and Rayleigh fading parameters of the distributions used to encode the $\mathcal{F}$ distribution, we first generated a set of $\mathcal{F}$ random variables and used the maximum likelihood estimation (MLE). The corresponding parameter estimates are also presented in Table~I. Interestingly, irrespective of the multipath fading ($m$) and shadowing ($m_s$) conditions, the estimated $\bar{\gamma}_R$ and $\bar{\gamma}_N$ are the same as the $\bar{\gamma}$. Consequently, the cross entropy between the $\mathcal{F}$ and Rayleigh distributions, i.e., \eqref{eqn35}, is dependent upon the average SNR only. When comparing the cross entropy for Rayleigh and Nakagami-$m$, for all of the cases, the Nakagami-$m$ distribution provided lower entropy than the Rayleigh distribution.

\begin{table}[!t]
\small
\renewcommand{\arraystretch}{1.3}
  \centering
  \caption{Shannon entropy and cross entropy for different fading parameters ($m$, $m_s$) and average SNR ($\bar{\gamma}$) along with the corresponding parameter estimates of the Rayleigh and Nakagami-$m$ distributions}
	\vspace{-0.6cm}
    \begin{tabularx}{0.85\textwidth}{c c X X X X c}
		\hline\hline
    \multirow{2}{*}{($m, m_s, \bar{\gamma}$)} & \multirow{2}{*}{$H(p)$} & \multicolumn{2}{c}{Rayleigh} & \multicolumn{3}{c}{Nakagami-$m$} \\
            &       &  \centering{$\bar{\gamma}_R$} &  \centering{$\!\!H(p,q)$} & \centering{$m$} &  \centering{$\bar{\gamma}_N$} &  $H(p,q)$ \\
				
		\hline
    (2, 3, 5 dB) & \centering{3.005}  & \centering{5 dB} & \centering{3.104} & \centering{1.14} & \centering{5 dB} & {3.096}  \\
    (2, 30, 5 dB) & \centering{2.959}  & \centering{5 dB} & \centering{3.104} & \centering{1.89} & \centering{5 dB} & 2.960  \\
    (20, 3, 5 dB) & \centering{2.730}  & \centering{5 dB} & \centering{3.104} & \centering{2.11} & \centering{5 dB} & 2.913  \\
    (20, 30, 5 dB) & \centering{1.870}  & \centering{5 dB} & \centering{3.104} & \centering{11.99} & \centering{5 dB} & 1.876  \\
		\hline
    (2, 3, 15 dB) & \centering{6.327}  & \centering{15 dB} & \centering{6.426} & \centering{1.14} & \centering{15 dB} & 6.418  \\
    (2, 30, 15 dB) & \centering{6.281}  & \centering{15 dB} & \centering{6.426} & \centering{1.88} & \centering{15 dB} & 6.282  \\
    (20, 3, 15 dB) & \centering{6.051}  & \centering{15 dB} & \centering{6.426} & \centering{2.11} & \centering{15 dB} & 6.235 \\
    (20, 30, 15 dB) & \centering{5.191}  & \centering{15 dB}  & \centering{6.426} & \centering{11.98} & \centering{15 dB} & 5.198  \\
    \hline\hline
    \end{tabularx}%
		\vspace{-0.8cm}
  \label{tab:addlabel}%
\end{table}%

From Table~I, it is also evident that the Shannon entropy was smaller than the cross entropy for all of the considered cases. Overall this  demonstrates the importance of considering composite fading models when characterization wireless transmission in conventional and emerging communication systems. It is worth remarking that for the light shadowing conditions (e.g., $m_s = 30$),  the cross entropy for Nakagami-$m$ distribution is almost the same as the Shannon entropy. This is due to the fact that the $\mathcal{F}$ distribution coincides with the Nakagami-$m$ distribution when $m_s \rightarrow \infty$. Consequently, its relative entropy\footnote{The relative entropy is given by $D(p||q) \triangleq  H(p,q) - H(p)$ \cite{kullback}.} (also known as the Kullback-Leibler divergence), which is a measure of the distance between two distributions, was close to zero.

\section{Conclusion}
		\vspace{-0.2cm}
In this paper, a comprehensive performance analysis of ED-based spectrum sensing over $\mathcal{F}$ composite fading channels has been carried out. A novel analytic expression for the average energy detection probability was derived and then extended to account for collaborative spectrum sensing, SLS diversity reception and noise power uncertainty. Additionally, as a  figure of merit to determine the performance of ED-based spectrum sensing, a closed-form expression for the AUC was also derived. It was shown that the detection performance increased when the average SNR increased, the time-bandwidth product decreased, or when the multipath fading and shadowing severity decreased. As anticipated, the detection performance was significantly improved as the number of diversity branches increased. Furthermore, when more collaborative users shared their local decision information, a better detection performance was achieved for both the OR- and AND-rules. Among these rules, the OR-rule was observed to provide a better detection performance compared to the AND-rule. To validate the analytical expressions presented in the paper, simulation results were also presented. 

Most importantly though, it is noted that the analytical form of the average detection probability for ED-based spectrum sensing over the generalized $K$ fading channels given in \cite[eq. (7)]{13} is only valid for integer value of $m$. However the analytical expression presented in the paper is valid for any $m$ value meaning that ED-based spectrum sensing may now be tested over a much greater range of multipath fading conditions, which is essential in demanding scenarios such as ED-based spectrum sensing and RADAR systems. Additionally, the analytical expression presented in this paper shows much less complexity due to the computation of a smaller number of special functions and a rapidly converging infinite series.

Novel expressions for the Shannon entropy and cross entropy were also derived in closed-form. The behavior of the Shannon entropy was evaluated for different values of the key parameters of $\mathcal{F}$ composite fading channels and then compared with the behavior of ROC and AUC, offering useful insights on the relationship of these measures. It was shown that the more bits were required to encode the corresponding message when the channel was subject to higher average SNR, severer mutlipath fading and heavier shadowing. Moreover, the cross entropy with the Rayleigh and Nakagami-$m$ distributions demonstrated the information loss encountered when the composite fading was not taken into account.

\vspace{-0.4cm}
\section*{Acknowledgement}
This work was supported in part by the U.K. Engineering and Physical Sciences Research Council under Grant No. EP/L026074/1 and the Department for the Economy Northern Ireland through Grant No. USI080.

\balance
\vspace{-0.7cm}
\bibliography{Manuscript}

\begin{thebibliography}{10}
\providecommand{\url}[1]{#1}
\csname url@samestyle\endcsname
\providecommand{\newblock}{\relax}
\providecommand{\bibinfo}[2]{#2}
\providecommand{\BIBentrySTDinterwordspacing}{\spaceskip=0pt\relax}
\providecommand{\BIBentryALTinterwordstretchfactor}{4}
\providecommand{\BIBentryALTinterwordspacing}{\spaceskip=\fontdimen2\font plus
\BIBentryALTinterwordstretchfactor\fontdimen3\font minus
  \fontdimen4\font\relax}
\providecommand{\BIBforeignlanguage}[2]{{%
\expandafter\ifx\csname l@#1\endcsname\relax
\typeout{** WARNING: IEEEtran.bst: No hyphenation pattern has been}%
\typeout{** loaded for the language `#1'. Using the pattern for}%
\typeout{** the default language instead.}%
\else
\language=\csname l@#1\endcsname
\fi
#2}}
\providecommand{\BIBdecl}{\relax}
\BIBdecl

\bibitem{1}
S.~Atapattu, C.~Tellambura, and H.~Jiang, ``Performance of an energy detector
  over channels with both multipath fading and shadowing,'' \emph{IEEE Trans.
  Wireless Commun.}, vol.~9, no.~12, pp. 3662--3670, Dec. 2010.

\bibitem{R1}
S.~Koteshwara and K.~K. Parhi, ``Incremental-precision based feature
  computation and multi-level classification for low-energy
  internet-of-things,'' \emph{IEEE J. Emerg. Sel. Topic Circuits Syst (Early
  Access)}, 2018.

\bibitem{R2}
Z.~Li, B.~Chang, S.~Wang, A.~Liu, F.~Zeng, and G.~Luo, ``Dynamic compressive
  wide-band spectrum sensing based on channel energy reconstruction in
  cognitive internet of things,'' \emph{IEEE Trans. Ind. Informat.}, vol.~14,
  no.~6, pp. 2598--2607, Jun. 2018.

\bibitem{R3}
L.~Yi, X.~Deng, M.~Wang, D.~Ding, and Y.~Wang, ``Localized confident
  information coverage hole detection in {I}nternet of things for radioactive
  pollution monitoring,'' \emph{IEEE Access}, vol.~5, pp. 18\,665--18\,674,
  Sep. 2017.

\bibitem{new1}
S.~Shukla, A.~K. Rao, and N.~Srivastava, ``A survey on energy detection schemes
  in cognitive radios,'' in \emph{Proc. ICETEESES}, Mar. 2016, pp. 223--228.

\bibitem{new2}
M.~Sardana and A.~Vohra, ``Analysis of different spectrum sensing techniques,''
  in \emph{Proc. Comptelix}, Jul. 2017, pp. 422--425.

\bibitem{new3}
P.~D. Sutton, K.~E. Nolan, and L.~E. Doyle, ``Cyclostationary signatures in
  practical cognitive radio applications,'' \emph{IEEE J. Sel. Areas Commun.},
  vol.~26, no.~1, pp. 13--24, Jan. 2008.

\bibitem{2}
H.~Urkowitz, ``Energy detection of unknown deterministic signals,'' \emph{Proc.
  IEEE}, vol.~55, no.~4, pp. 523--531, Apr. 1967.

\bibitem{3}
A.~Ghasemi and E.~S. Sousa, ``Spectrum sensing in cognitive radio networks:
  {R}equirements, challenges and design trade-offs,'' \emph{IEEE Commun. Mag.},
  vol.~46, no.~4, Apr. 2008.

\bibitem{4}
Y.-C. Liang, K.-C. Chen, G.~Y. Li, and P.~Mahonen, ``Cognitive radio networking
  and communications: An overview,'' \emph{IEEE Trans. Veh. Technol.}, vol.~60,
  no.~7, pp. 3386--3407, Sep. 2011.

\bibitem{5}
F.~F. Digham, M.-S. Alouini, and M.~K. Simon, ``On the energy detection of
  unknown signals over fading channels,'' in \emph{Proc. IEEE ICC}, May 2003,
  pp. 3575--3579.

\bibitem{6}
F.~B.~S. de~Carvalho, W.~T.~A. Lopes, and M.~S. Alencar, ``Performance of
  cognitive spectrum sensing based on energy detector in fading channels,''
  \emph{Procedia Comput. Sci.}, vol.~65, pp. 140--147, 2015.

\bibitem{7}
D.~Horgan and C.~C. Murphy, ``Fast and accurate approximations for the analysis
  of energy detection in {N}akagami-$m$ channels,'' \emph{IEEE Commun. Lett.},
  vol.~17, no.~1, pp. 83--86, Jan. 2013.

\bibitem{8}
S.~Hussain and X.~N. Fernando, ``Closed-form analysis of relay-based cognitive
  radio networks over {N}akagami-$m$ fading channels,'' \emph{IEEE Trans. Veh.
  Technol.}, vol.~63, no.~3, pp. 1193--1203, Sep. 2014.

\bibitem{9}
F.~F. Digham, M.-S. Alouini, and M.~K. Simon, ``On the energy detection of
  unknown signals over fading channels,'' \emph{IEEE Trans. Commun.}, vol.~55,
  no.~1, pp. 21--24, Jan. 2007.

\bibitem{R4}
P.~C. Sofotasios, A.~Bagheri, T.~A. Tsiftsis, S.~Freear, A.~Shahzadi, and
  M.~Valkama, ``A comprehensive framework for spectrum sensing in non-linear
  and generalized fading conditions,'' \emph{IEEE Trans. Veh. Technol},
  vol.~66, no.~10, pp. 8615--8631, Oct. 2017.

\bibitem{10}
H.~Rasheed, F.~Haroon, and N.~Rajatheva, ``Performance analysis of
  {R}ice-lognormal channel model for spectrum sensing,'' in \emph{Proc.
  ECTI-CON}, May 2010, pp. 420--424.

\bibitem{11}
S.~Atapattu, C.~Tellambura, and H.~Jiang, ``Energy detection based cooperative
  spectrum sensing in cognitive radio networks,'' \emph{IEEE Trans. Wireless
  Commun.}, vol.~10, no.~4, pp. 1232--1241, Apr. 2011.

\bibitem{12}
N.~Reisi, S.~Gazor, and M.~Ahmadian, ``Distributed cooperative spectrum sensing
  in mixture of large and small scale fading channels,'' \emph{IEEE Trans.
  Wireless Commun.}, vol.~12, no.~11, pp. 5406--5412, Nov. 2013.

\bibitem{13}
S.~Atapattu, C.~Tellambura, and H.~Jiang, ``Performance of an energy detector
  over channels with both multipath fading and shadowing,'' \emph{IEEE Trans.
  Wireless Commun.}, vol.~9, no.~12, pp. 3662--3670, Dec. 2010.

\bibitem{14}
K.~P. Peppas, G.~Efthymoglou, V.~A. Aalo, M.~Alwakeel, and S.~Alwakeel,
  ``Energy detection of unknown signals in {G}amma-shadowed {R}ician fading
  environments with diversity reception,'' \emph{IET Commun.}, vol.~9, no.~2,
  pp. 196--210, Jan. 2014.

\bibitem{15}
H.~Al-Hmood and H.~S. Al-Raweshidy, ``Unified modeling of composite
  $\kappa$-$\mu$ / gamma, $\eta$-$\mu$ / gamma and $\alpha$-$\mu$ / gamma
  fading channels using a mixture gamma distribution with applications to
  energy detection,'' \emph{IEEE Antennas and Wireless Propag. Lett.}, vol.~16,
  pp. 104--108, Feb. 2017.

\bibitem{16}
S.~K. Yoo, S.~L. Cotton, P.~C. Sofotasios, M.~Matthaiou, M.~Valkama, and G.~K.
  Karagiannidis, ``The {F}isher-{S}nedecor $\mathcal{F}$ distribution: A simple
  and accurate composite fading model,'' \emph{IEEE Commun. Lett.}, vol.~21,
  no.~7, pp. 1661--1664, Jul. 2017.

\bibitem{entropy1}
Y.-Y. Zhu and Y.-G. Zhu, ``The simulation study of entropy-based signal
  detector over fading channel,'' in \emph{Proc. WCSP}, Oct. 2012, pp. 1--5.

\bibitem{17}
I.~S. Gradshteyn and I.~M. Ryzhik, \emph{Table of Integrals, Series, and
  Products}, 7th~ed.\hskip 1em plus 0.5em minus 0.4em\relax London: Academic
  Press, 2007.

\bibitem{18}
M.~Abramowitz and I.~A. Stegun, \emph{Handbook of mathematical
  functions}.\hskip 1em plus 0.5em minus 0.4em\relax Washington, DC. US Dept.
  of Commerce, National Bureau of Standards, 1972.

\bibitem{19}
P.~E. Cantrell and A.~K. Ojha, ``Comparison of generalized {Q}-function
  algorithms,'' \emph{IEEE Trans. Inf. Theory}, vol.~33, no.~4, pp. 591--596,
  Jul. 1987.

\bibitem{20}
V.~M. Kapinas, S.~K. Mihos, and G.~K. Karagiannidis, ``On the monotonicity of
  the generalized {M}arcum and {N}uttall {Q}-functions,'' \emph{IEEE Trans.
  Inf. Theory}, vol.~55, no.~8, pp. 3701--3710, Aug. 2009.

\bibitem{21}
Wolfram Research, Inc., 2018, visited on 07/13/2018. [Online]. Available:
  \url{http://functions.wolfram.com/id}.

\bibitem{23}
A.~Ghasemi and E.~S. Sousa, ``Impact of user collaboration on the performance
  of sensing-based opportunistic spectrum access,'' in \emph{Proc. IEEE VTC},
  Sep. 2006, pp. 1--6.

\bibitem{24}
D.~Duan, L.~Yang, and J.~C. Principe, ``Cooperative diversity of spectrum
  sensing for cognitive radio systems,'' \emph{IEEE Trans. Signal Process.},
  vol.~58, no.~6, pp. 3218--3227, Jun. 2010.

\bibitem{25}
R.~Tandra and A.~Sahai, ``{SNR} walls for signal detection,'' \emph{IEEE J.
  Sel. Topics Signal Process.}, vol.~2, no.~1, pp. 4--17, Feb. 2008.

\bibitem{26}
Y.~Xiao and F.~Hu, \emph{Cognitive radio networks}.\hskip 1em plus 0.5em minus
  0.4em\relax CRC press, 2008.

\bibitem{R5}
S.~Dikmese, Z.~Ilyas, P.~C. Sofotasios, M.~Renfors, and M.~Valkama, ``Sparse
  frequency domain spectrum sensing and sharing based on cyclic prefix
  autocorrelation,'' \emph{IEEE J. Sel. Areas Commun.}, vol.~35, no.~1, pp.
  159--172, Jan. 2017.

\bibitem{R6}
S.~Dikmese, P.~C. Sofotasios, M.~Renfors, and M.~Valkama, ``Subband energy
  based reduced complexity spectrum sensing under noise uncertainty and
  frequency-selective spectral characteristics,'' \emph{IEEE Trans. Signal
  Process.}, vol.~64, no.~1, pp. 131--145, Jan. 2016.

\bibitem{27}
M.~Lopez-Benitez and F.~Casadevall, ``Signal uncertainty in spectrum sensing
  for cognitive radio,'' \emph{IEEE Trans. Commun.}, vol.~61, no.~4, pp.
  1231--1241, Apr. 2013.

\bibitem{28}
T.~D. Wickens, \emph{Elementary signal detection theory}.\hskip 1em plus 0.5em
  minus 0.4em\relax Oxford University Press, USA, 2002.

\bibitem{29}
J.~A. Hanley and B.~J. McNeil, ``The meaning and use of the area under a
  receiver operating characteristic ({ROC}) curve,'' \emph{Radiology}, vol.
  143, no.~1, pp. 29--36, Apr. 1982.

\bibitem{R7}
S.~Dikmese, P.~C. Sofotasios, T.~Ihalainen, M.~Renfors, and M.~Valkama,
  ``Efficient energy detection methods for spectrum sensing under non-flat
  spectral characteristics,'' \emph{IEEE J. Sel. Areas Commun.}, vol.~33,
  no.~5, pp. 755--770, May 2015.

\bibitem{30}
S.~Atapattu, C.~Tellambura, and H.~Jiang, ``Analysis of area under the {ROC}
  curve of energy detection,'' \emph{IEEE Trans. Wireless Commun.}, vol.~9,
  no.~3, Mar. 2010.

\bibitem{31}
T.~Fawcett, ``An introduction to {ROC} analysis,'' \emph{Pattern Recognit.
  Lett.}, vol.~27, no.~8, pp. 861--874, Jun. 2006.

\bibitem{new_AUC}
S.~Alam, O.~Olabiyi, O.~Odejide, and A.~Annamalai, ``A performance study of
  energy detection for dual-hop transmission with fixed gain relays: area under
  {ROC} curve {(AUC)} approach,'' in \emph{Proc. PIMRC}, Sep. 2011, pp.
  1840--1844.

\bibitem{entropy2}
P.~C. Sofotasios, S.~Muhaidat, M.~Valkama, M.~Ghogho, and G.~K. Karagiannidis,
  ``Entropy and channel capacity under optimum power and rate adaptation over
  generalized fading conditions,'' \emph{IEEE Signal Process. Lett.}, vol.~22,
  no.~11, pp. 2162--2166, Nov. 2015.

\bibitem{kullback}
S.~Kullback, \emph{Information theory and statistics}.\hskip 1em plus 0.5em
  minus 0.4em\relax Courier Corporation, 1997.

\end{thebibliography}
\bibliographystyle{IEEEtran}

\end{document}